\documentclass{aastex631}

\graphicspath{{./}{figures/}}

\begin{document}


\title{Deep {\em Chandra} Observation of the Remarkable Ionization Cones of NGC 5252}

\correspondingauthor{J. Wang}
\email{jfwang@xmu.edu.cn}

\author{Chen Wang}
\affiliation{Department of Astronomy, Xiamen University, Xiamen, 361005, China}
\affiliation{Universit$\acute{\rm e}$ Paris Saclay and Universit$\acute{\rm e}$ Paris Cit$\acute{\rm e}$, CEA, CNRS, AIM, F-91191 Gif-sur-Yvette Cedex, France}

\author{Junfeng Wang}
\affiliation{Department of Astronomy, Xiamen University, Xiamen, 361005, China}

\author{Mauro Dadina}
\affiliation{INAF/IASF Bologna via Gobetti 101, 40129, Bologna, Italy}

\author{Giuseppina Fabbiano}
\affiliation{Harvard-Smithsonian Center for Astrophysics, 60 Garden St, Cambridge, MA 02138, USA}

\author{Martin Elvis}
\affiliation{Harvard-Smithsonian Center for Astrophysics, 60 Garden St, Cambridge, MA 02138, USA}

\author{Stefano Bianchi}
\affiliation{Universit$\grave{\rm a}$ degli Studi Roma Tre, via della Vasca Navale 84, I-00146 Roma, Italy}

\author{Matteo Guainazzi}
\affiliation{
ESA European Space Research and Technology Centre (ESTEC), Keplerlaan 1, 2201 AZ, Noordwijk, Netherland}




\begin{abstract}

Seyfert galaxy NGC 5252 harbors enormously extended ionization cones, which have been previously detected in the optical and X-ray band, offering a unique opportunity to investigate the interaction between the central active galactic nucleus (AGN) and the surrounding gas in the AGN host galaxy. We present deep {\em Chandra} imaging spectroscopy of NGC 5252 with a total exposure time of 230 ks. The morphology in the soft X-rays shows resolved extended structure from nucleus to a large radial distance, and for the first time we detect the outermost X-ray arc at $\sim$20 kpc. The X-ray cone mostly follows the direction of the optical ionization cones in the southeast and northwest direction, about 20 degrees misaligned with the major axis of the galactic disk of NGC 5252. Fitting the spectra extracted from radial sectors with photoionization models supports that extended emission is mainly photoionized by the central AGN. We also examine the variation of the photoionization parameter along the radial extension, and infer a decreasing ionizing continuum of the central engine by a factor of $\sim$50 over the past 64000 years. These findings are consistent with previous suggestion that NGC 5252 resembles a quasar relic with a $M\sim 10^9 M_{\odot}$ supermassive black hole, which went through a minor merger event and switched to a low accretion rate state.

\end{abstract}

\keywords{X-rays: galaxies --- galaxies: emission line --- galaxies: ISM --- galaxies: individual (NGC 5252)}


\section{Introduction} \label{sec:intro}

The conical structure extending to kilo-parsec scale, known as ''ionization cones'' \citep{1989Natur.341..422T}, is almost ubiquitously seen in extensive optical emission line studies of narrow-line regions (NLRs) in nearby Seyfert galaxies \citep{2003ApJ...597..768S}.  Such a large scale anisotropic emission feature is among the best evidences for the Active Galactic Nuclei (AGNs) unification model, interpreted as gas ionized by the radiation escaping the AGN collimated by the putative torus \citep{1996ApJ...467..197M,2015ARA&A..53..365N}. In terms of understanding the kinetic and radiative output from AGN, such extended emission line regions uniquely provide a direct view of AGN--host galaxy interaction \citep[e.g.,][]{1992ApJ...396...45S,1999MNRAS.309....1F}. 

One of the most extended ionization cones ($\sim$40 kpc across) in the optical ever observed in nearby galaxies is harbored in NGC 5252 \citep{1989Natur.341..422T}, an early type (S0) Seyfert 1.9 galaxy \citep{1993ApJ...414..552O} at redshift $z$=0.023 (corresponding to a luminosity distance $D_L$=100.8 Mpc and a scale of 0.467 kpc per arcsec; NED\footnote{\url{http://ned.ipac.caltech.edu/}}). Images of [OIII] line emission obtained by \citet{1989Natur.341..422T} revealed several arcs confined in a well-defined bi-cone encircling the nucleus of the galaxy. The outermost arcs extend to a radius of 50 arcsec ($\sim$ 23 kpc) to the north of the nucleus and 45 arcsec ($\sim$ 20 kpc) to the south.

\citet{1998ApJ...505..159M} studied the fine scale morphology of the extended gas with {\em Hubble Space Telescope} (HST)/WFPC2 and further identified three dynamical components in gas kinematics by mapping the velocity field of the ionized gas.  The significant misalignment between one circumnuclear gas disk and the stellar disk leads to the conclusion that NGC 5252 possibly went through a galaxy merger \citep{1998ApJ...505..159M}. The lack of close companion galaxies to NGC 5252 \citep{1996A&A...306...39F} and absence of disturbed velocity field of the gas indicate that it is unlikely that the merger event occurred in the recent past, instead the observed emission line arcs are likely part of the debris gas from this putative galaxy-galaxy encounter illuminated by the AGN \citep{1998ApJ...505..159M}.

VLA observations revealed the arcsecond scale resolution radio structure of NGC 5252 \citep{1994AJ....107.1227W}, which contains an unresolved core with position corresponding to the optical nucleus. Weak jet-like extensions to the north and south of the nucleus were also resolved, with position angles (PAs) of $345\pm3$ degree and $175\pm5$ degree, respectively. An off-nuclear radio source $22^{\prime\prime}$ north of the nucleus was later identified with an ultra-luminous X-ray source (ULX) and multi-band followup studies suggest an accreting intermediate mass black hole ($>10^4$M$\odot$) which might be the stripped remnant of a merging dwarf galaxy~\citep{2015ApJ...814....8K,2017ApJ...844L..21K}.  \citet{2017MNRAS.464L..70Y} performed VLBI observation of the off-nuclear radio source, and further confirmed that the ULX presents the similar characteristics of a weakly accreting supermassive black hole (SMBH). Therefore NGC 5252 may consist of a relatively rare pair of radio-emitting AGNs, also consistent with having experienced a merger in the past. Furthermore it is worth noting that the mass of the central SMBH of NGC 5252 measured with {\em HST}/STIS long-slit spectroscopy is $M_{BH}=0.95^{+1.45}_{-0.45}\times 10^9 M_{\odot}$, substantially larger than that of a typical Seyfert galaxy \citep{2005A&A...431..465C}. The hard X-ray luminosity implies the nucleus is emitting much below its Eddington limit \citep[0.005 $L_{Edd}$;][]{2005A&A...431..465C}, consistent with an observed flat power-law X-ray spectrum of NGC 5252 \citep[$\Gamma \sim 1.4-1.5$;][]{1996ApJ...456..141C,2010A&A...516A...9D}. 

Altogether these pieces of evidence raised an intriguing conjecture that the active nucleus of NGC 5252 may be a quasar ``relic'' \citep{2005A&A...431..465C}. Indeed, the large-scale extended ionized gas around the galaxies could contain the chronicle of the past AGN accretion activities due to the light travel time: gaseous debris located at different radial distances were illuminated by ionizing photons emitted from the central source at different time. For instance, extended emission line clouds called ``Hanny's Voorwerp'' have been identified, where the circumambient gas appears to be ionized by a luminous quasar whereas the nucleus somehow had faded significantly to a low luminosity AGN \citep{2009MNRAS.399..129L}. Further investigations show that the majority of AGNs in a sample of galaxies which are encompassed by AGN-ionized clouds had faded over a time scale of tens of thousands years \citep{2012MNRAS.420..878K}. Not surprisingly, NGC 5252 belongs to this newly identified class of Hanny's Voorwerp \citep{2015AJ....149..155K}, and the fact that it contains one of the most extended cones serves as a great opportunity to reconstruct its SMBH activities during a long timespan \citep{2012MNRAS.420..878K}.

Previous observations with {\em XMM}-Newton and {\em Chandra} have examined the extended X-ray emission in this context \citep{2010A&A...516A...9D}. The spatially resolved soft X-ray emission around the nucleus is attributed to ionized gas and well overlaps with the optical structures observed in the [OIII] $\lambda 5007$\AA\/ emission line. \citet{2010A&A...516A...9D} further analyzed the flux ratios of [OIII] and soft X-rays in the extended zones from about 100 pc to 1.5 kpc, and suggest that the source is indeed a quasar relic with radiatively inefficient accretion in a low state.  

The {\em Chandra} exposure of NGC 5252 used in previous work was relatively short ($\sim$ 60 ks) for detecting X-ray emission in the larger scale extension. Since the [O III] image obtained by \citet{1989Natur.341..422T} reveals a much larger extension of ionized gas, it is expected that X-ray emission of lower surface brightness exists in outer regions as well \citep{2006A&A...448..499B}. Moreover, with improved statistics for spectral analysis, the structures at different radial distance to the nucleus may provide more details of the luminosity variation behavior of the ionizing nucleus, probing beyond the timespan covered by \citet{2010A&A...516A...9D}. This motivated us to obtain deeper exposure of this archetypal Seyfert galaxy with ionization cones. With a total exposure time of 230 ks, the full data set of {\em Chandra} observations used in this work is nearly four times deeper than the previous one, enabling us to investigate the spatially resolved properties of the extended emission to a large radial distance from the nucleus. In this work we focus on the extended emission, and the X-ray spectral analysis of the point sources including the nucleus will be reported separately (C. Wang et al. in preparation). The structure of this paper is as follows. Section 2 describes the observation and data reduction. In Section 3 we present the imaging and spectral properties of the extensions, and discuss the results in Section 4. Lastly Section 5 summarizes the findings of our study. 

\section{Observations and Data Reduction}

The {\em Chandra} X-ray observatory observed NGC 5252 four times. The first exposure was in 2003 and the other three were in 2013. All observations were made by the Advanced CCD Imaging Spectrometer (\emph{ACIS-S}; Garmire et al. 2003) in ``Very Faint'' mode. The nucleus of NGC 5252 was placed near the aimpoint on the backside-illuminated S3 chip, which has good sensitivity to soft X-rays. Details of these four observations are listed in Table \ref{io}.  For bright X-ray sources like NGC 5252, photon ``pile-up'' may occur when more than one photons strike the detector in a single frame of exposure, and potentially recognized by the detector as events with bad grade. To reduce the pile-up effect, our observations were made in 1/4 sub-array configuration and the resulting ACIS frame time is 0.8 seconds.
    
The tools that we used to reduce the {\em Chandra} data are the software package {\em Chandra} Interactive Analysis of Observations (CIAO, version 4.8) with the {\em Chandra} Calibration Database (CALDB version 4.7.2) files  \citep{2006SPIE.6270E..60F} provided by {\em Chandra} X-ray Center (CXC). We followed the standard pipeline and procedures described in our previous work \citep[e.g.,][]{Wang11a,Wang14}. First, we reprocessed all level=1 event files with $\emph{chandra\_repro}$ to obtain the level=2 event files filtered on grade and status, and to update the observation specific auxiliary files like the bad pixel map. Light curves created from source free regions were analyzed to identify periods of anomalous background flares. The total good exposure times after filtering are 59.8 ks, 40.4 ks, 67.6 ks, and 62.3 ks for the four Obs-IDs sequentially. The CCD readout streaks have been removed. Table \ref{io} summarizes the observations used in this paper. The \emph{wavdetect} tool was run on the full band (0.3--8 keV) images in wavelet scales from 1 to 16 pixel in steps of $\sqrt{2}$ to search for point sources on the S3 chip, which were removed later for the diffuse emission analysis. Finally we merged all event files from the four observations and created exposure corrected images.  The merged {\em Chandra} ACIS images of the central $2\arcmin \times 2\arcmin$ ($\sim 56 \times 56$ kpc) region of NGC 5252 were extracted in the 0.3--2 keV and 2--8 keV energy bands, as shown in Figure~\ref{cmap}. Thanks to the deeper total exposure, these images show significantly more structure than the first ACIS image obtained by \citet{2010A&A...516A...9D}. The details will be described in the next session. 

The source spectra and instrument responses including ancillary response files (ARFs) and response matrix files (RMFs) were generated by CIAO tool {\em specextract}, and analyzed through {\em XSPEC} (version 12.8.2) \citep{1996ASPC..101...17A}. Spectra were grouped to have a minimum of 15 counts per energy bin so that $\chi^2$ statistics could be used for the fitting.  The background spectra were extracted from a region free of point sources in the same CCD. Spectral modeling was performed in the energy range between 0.3 keV and 8 keV, to avoid the instrument calibration uncertainty at lower energies and the higher background at high energies.

\section{Imaging and Spectral Analysis}

\subsection{Point-spread Function (PSF) of the Nucleus}

Since NGC 5252 hosts a bright X-ray nucleus, it is important to characterize the PSF of the point like nucleus to study the extended X-ray emission.  We need to evaluate the pile-up even though ACIS sub-array mode was already employed to mitigate this effect. This step was accomplished with CIAO tool {\em pileup\_map}\footnote{\url{http://cxc.harvard.edu/ciao/ahelp/pileup\_map.html}}, which estimates that the pile-up fraction in the central part of NGC 5252 is about $15\%$.  Spectrum of the nucleus was extracted from a circular region with a radius of 2 arcsec. In \citet{2010A&A...516A...9D}, the best-fit model for the NGC 5252 nuclear spectrum consists of two absorbed power law components with photon index $\Gamma$ of 1.05 and 3 using {\em XMM}-Newton observation not affected by pile-up. We fit the {\em Chandra} spectrum following this model with these two photon index fixed and with pile-up correction. A satisfactory fit was obtained ($\chi^2/dof=601.2/516$). We also attempted fitting with the index of power law components linked and set as a free parameter. The best-fit is obtained with $\Gamma=1.14$, which slightly improved the fit ($\chi^2/dof=576.9/515$). As mentioned above, details of fitting the nuclear spectrum with various complex spectral models will be reported elsewhere (C. Wang et al. in preparation). To check the variability of the nucleus of NGC 5252, we used the Gregory-Loredo algorithm with CIAO tool {\em gl$\_$vary}\footnote{\url{http://cxc.harvard.edu/ciao/threads/variable}}. The variability indices are all 0 which indicates the nucleus is not variable during all four observations. Additionally, the flux of the nucleus remains the same in the four exposures, $F_{0.5-8{\rm keV}}=1.0\pm0.1 \times 10^{-11} $ erg s$ ^{-1}$ cm$^{-2}$. This result indicates that the observed luminosity of NGC 5252 nucleus is stable at $L_{0.5-8{\rm keV}}= 1.2\pm0.1 \times 10^{43} $ erg s$ ^{-1}$  since the first observation. 


To make the PSF simulation of the nucleus, the intrinsic nuclear spectrum was uploaded to {\em Chandra} Ray Tracing (ChaRT\footnote{\url{http://cxc.harvard.edu/ciao/PSFs/chart2/index.html}}) for the generation of the simulated ray trace file, which was then transformed to events file through the {\tt Marx} simulator  \citep{2012SPIE.8443E..1AD} . Figure~\ref{fig1}a and~\ref{fig1}b compare the observed radial profiles of the nuclear region of NGC 5252 with that of the simulated PSF in the 0.3--2.0 keV and 2.0--8.0 keV band, respectively. The soft (0.3--2 keV) X-ray emission close to the nucleus is clearly resolved, with extended emission dominates over the nuclear PSF wings at radii $>2 \arcsec$ from the nucleus. There is no evidence for extended hard band (2.0--8.0 keV) emission from the radial profile, although there could be some within $2 \arcsec$ \citep[e.g., seen in other Seyfert galaxies;][]{2020ApJ...900..164M,2023ApJ...951..146M}. This is beyond the scope of current work on large scale ionization cone and will be further investigated in the study of the nucleus.

\subsection{Morphology of the X-ray Bi-cone}
\label{section:subcone}
\begin{figure}[ht]
\centering  
{
\includegraphics[angle=0,width=8.5cm,clip]{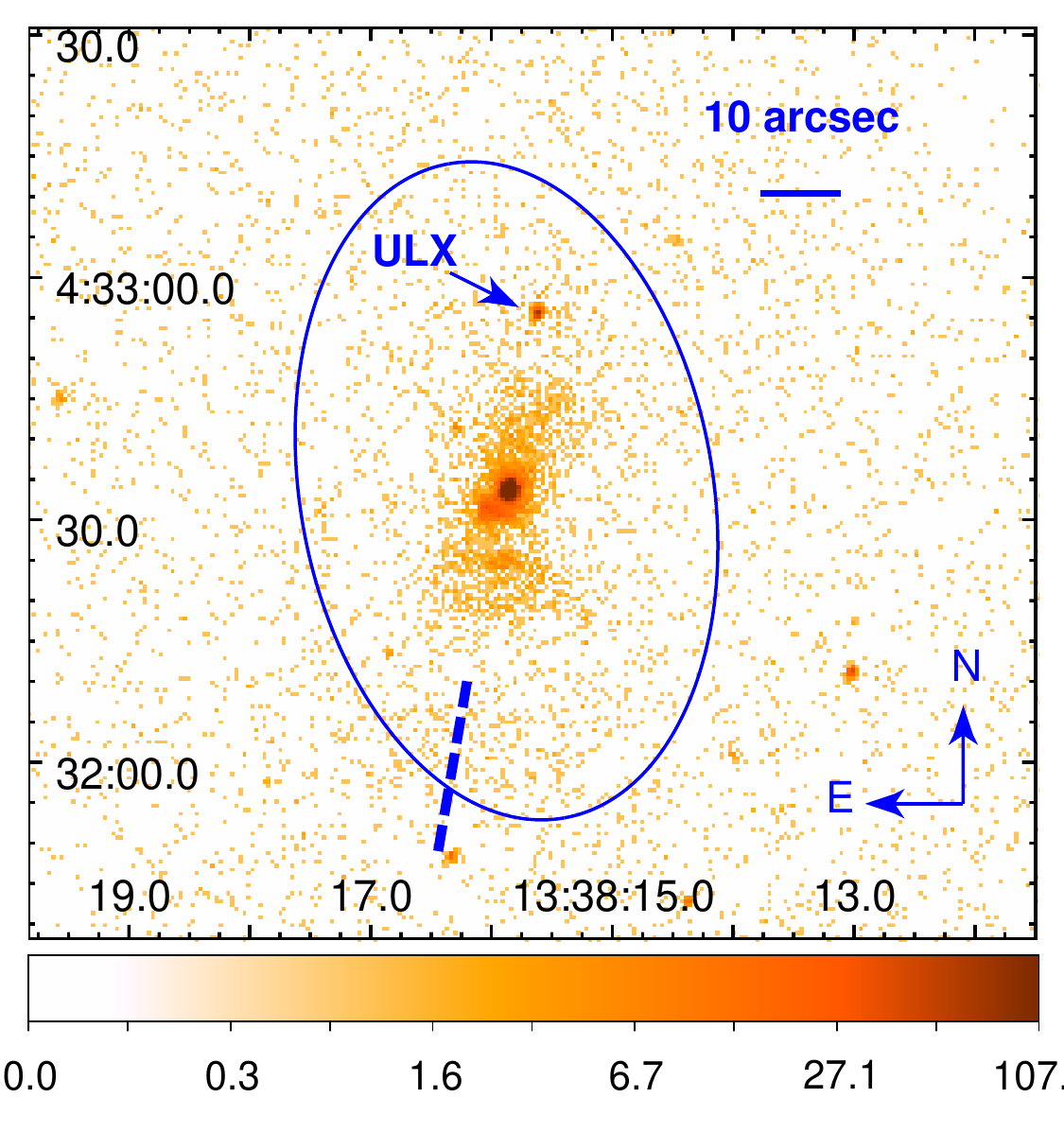}
\includegraphics[angle=0,width=8.5cm,clip]{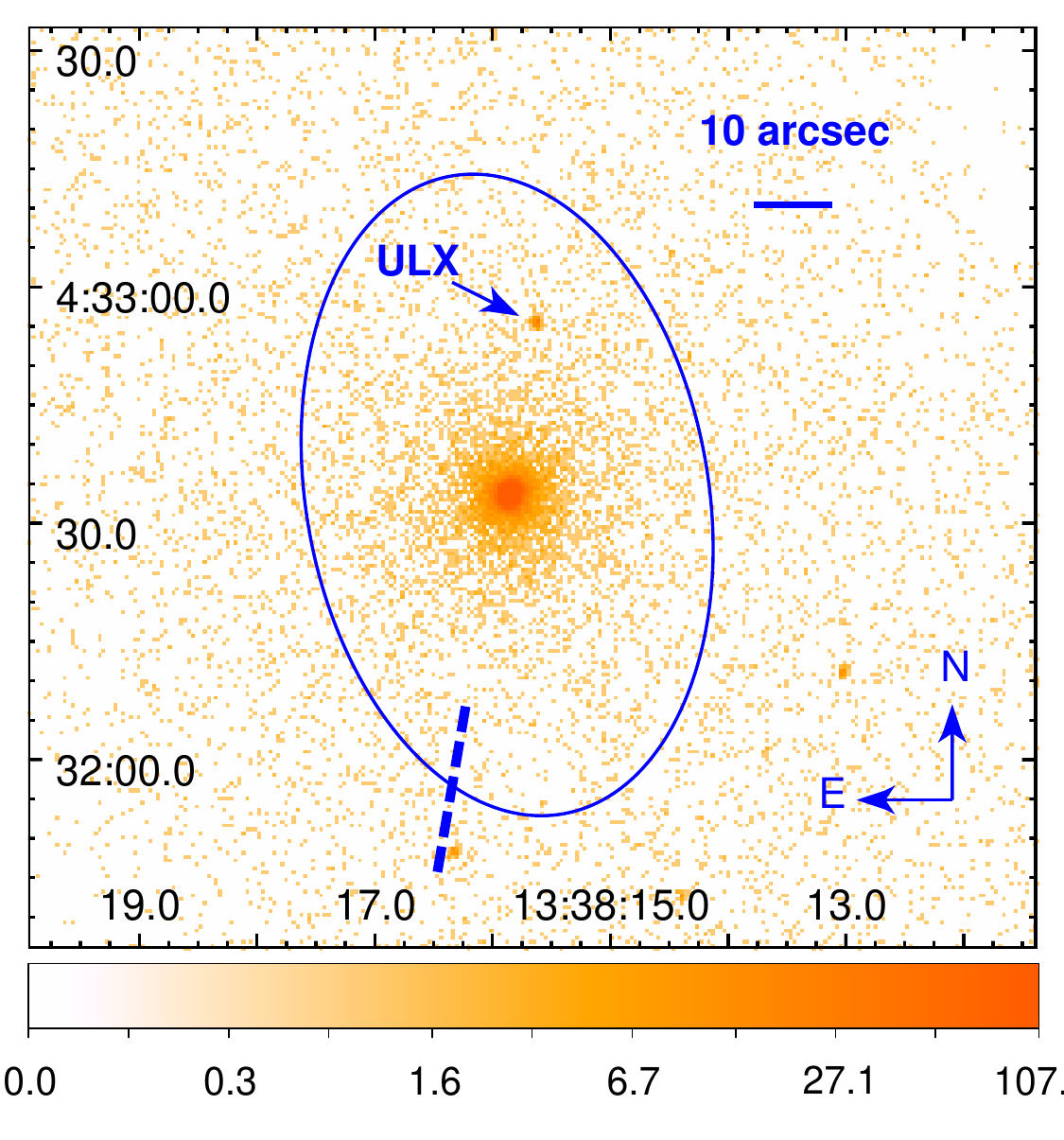}
\includegraphics[angle=0,width=10cm,clip]{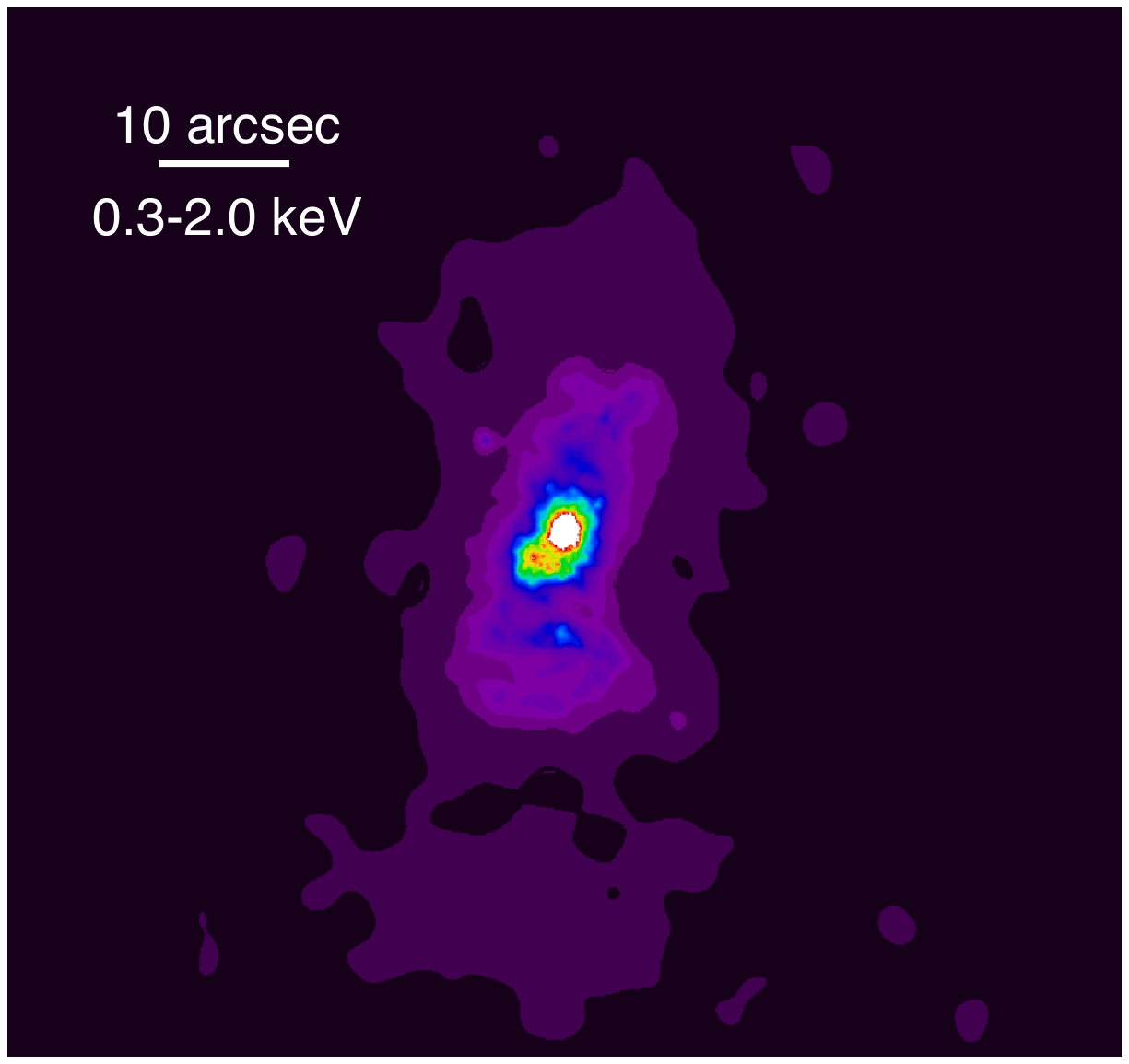}
}  
\caption{{\em Chandra} ACIS-S image of NGC 5252 in 0.3--2.0 keV (upper-left panel) and 2.0--8.0 keV (upper-right panel), extracted using merged data from the four {\em Chandra} observations listed in Table \ref{io}. The ULX is CXO J133815.6+043255. The dashed line shows the axis of the soft X-ray extensions. The ellipse exhibits the size and position of the host galaxy in 25-mag. To illustrate the faintest structure of the extended soft X-rays, the lower panel shows the adaptively smoothed {\em Chandra} ACIS-S image of NGC 5252 in 0.3--2.0 keV. Point sources have been removed from this image.}
\label{cmap}   
\end{figure}

\begin{figure}[ht]
\centering  
{  
\includegraphics[width=10cm, clip]{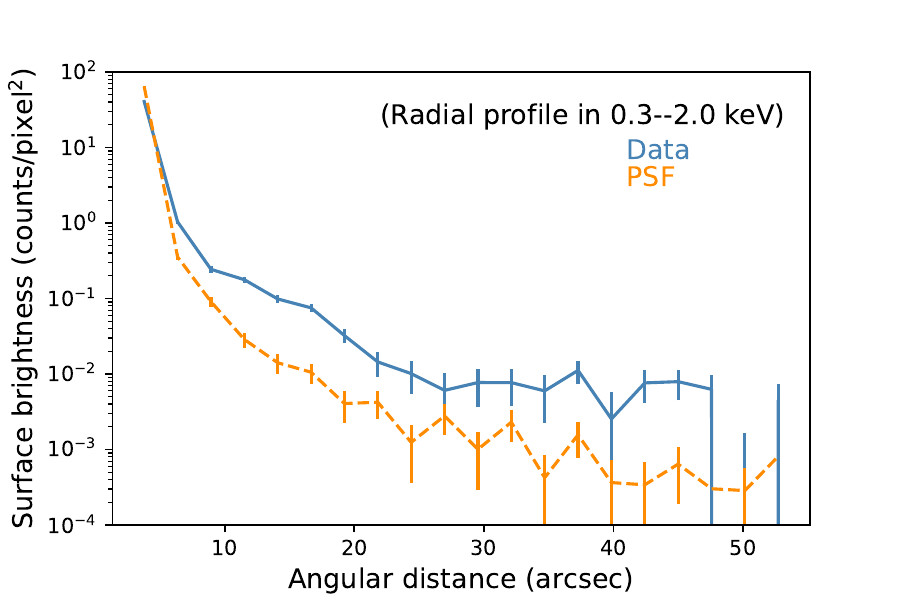}  
\includegraphics[width=10cm, clip]{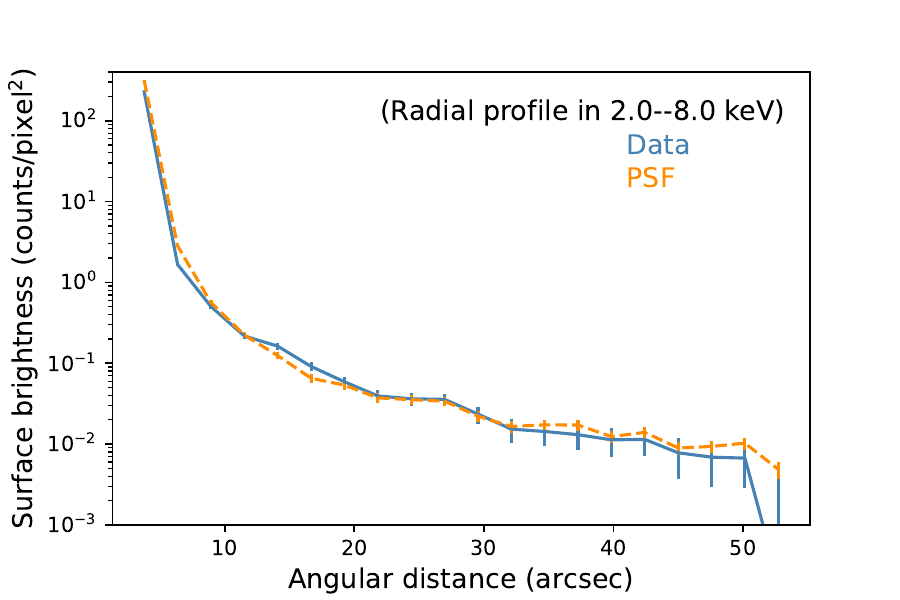}
}  
\caption{The radial profiles of NGC 5252 in two energy bands, 0.3--2.0 keV (upper panel) and 2.0--8.0 keV (lower panel). The dashed line shows the profile of the PSF simulation of the nucleus of NGC 5252. And the straight line is the profile of the observed data. The data shown here is from a single {\em Chandra} ObsID 15022. The excess over nuclear emission is only seen in the soft X-ray band.}
\label{fig1}  
\end{figure}

\begin{figure}[ht]
\centering  
{  
\includegraphics[angle=270,width=7cm,clip]{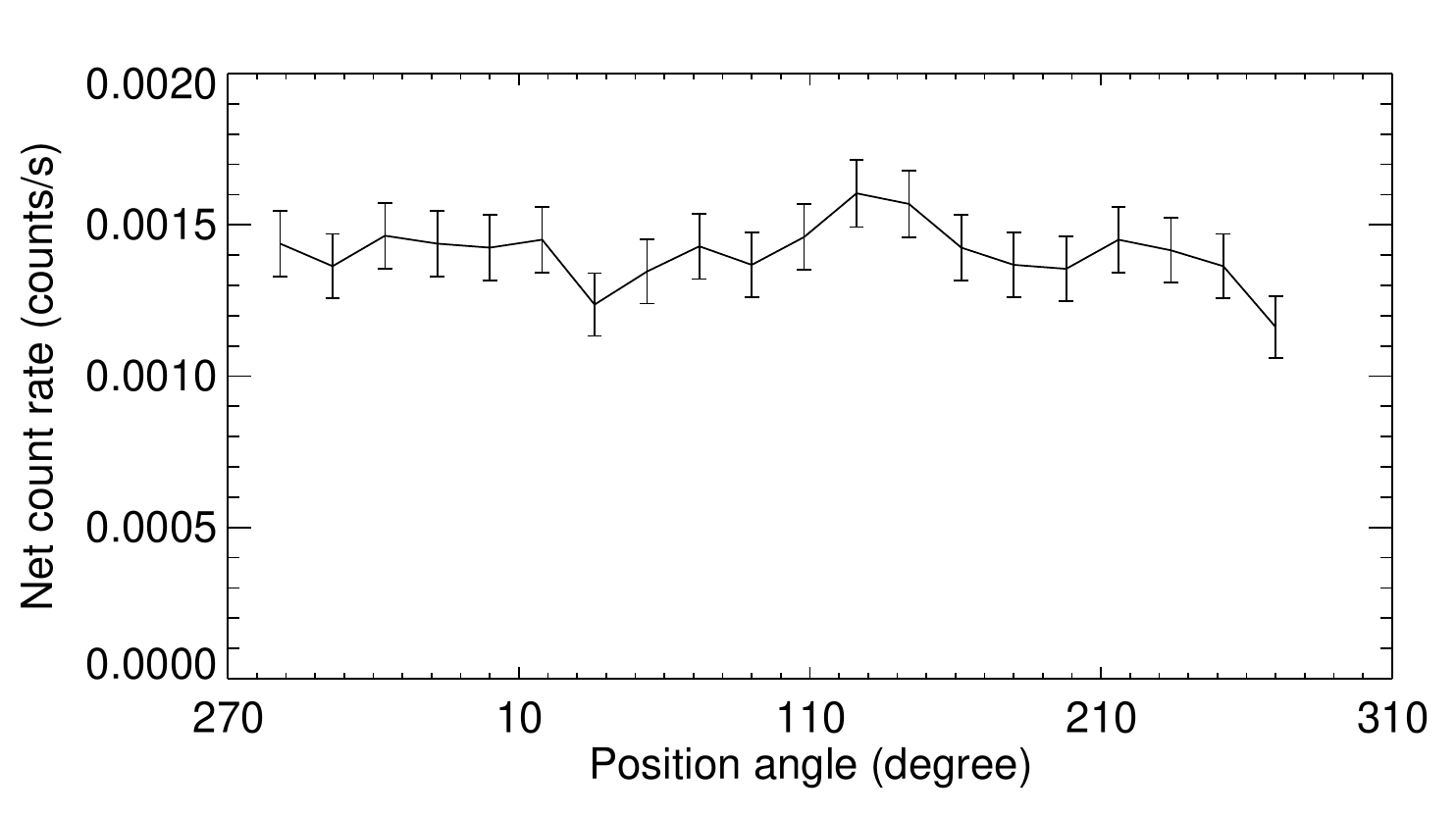} 
\includegraphics[angle=270,width=7cm,clip]{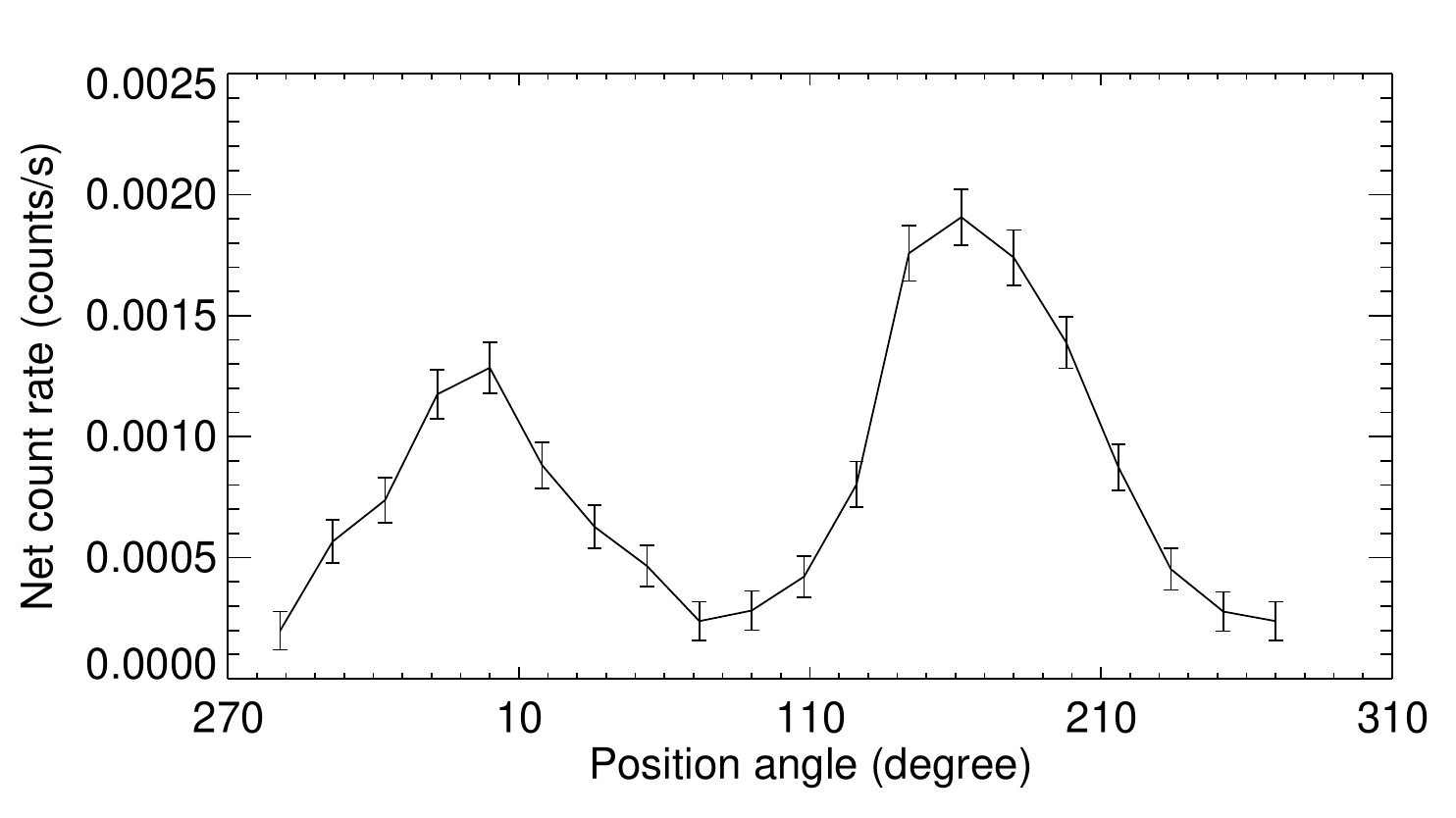}
}  
\caption{The azimuthal profiles of NGC 5252 in two energy bands. The merged data of four {\em Chandra} observations (see Table \ref{io}) is used. The left one depicts the profile in 2.0--8.0 keV, while the right one is the profile in 0.3--2.0 keV.}  
\label{com}  
\end{figure}

Figure~\ref{cmap}a and~\ref{cmap}b show the ACIS images of NGC 5252 extracted from merged event files in the 0.3--2.0 keV and 2.0--8.0 keV band, respectively.  The existence of a biconical morphology of the extended emission is apparent in the soft X-ray image (Figure~\ref{cmap}a) whilst no extended feature is seen in the hard band. In the soft band it is evident that the soft X-ray emission is concentrated in two cones, and the direction of the extension in the northwest (NW) is closely aligned to that in the southeast (SE).  To quantify the morphology of the extended bi-cones seen in Figure~\ref{cmap}a, we extracted counts from 20 azimuthal sectors centered on the nucleus with an apex angle of 18 degree, and obtained the azimuthal surface brightness profiles in 0.3--2.0 keV and 2.0--8.0 keV ($2\arcsec \leq r \leq 40\arcsec)$.  These azimuthal variations, shown in Figure~\ref{com}, clearly outline the sectors that contain the bright cones. The PA of the axis of NW sector and SE sector is $-10$ degree and 170 degree, respectively. There is no excess in 2--8 keV emission seen across the different azimuthal directions. 

\begin{figure}
\centering 
{  
\includegraphics[width=10cm,clip]{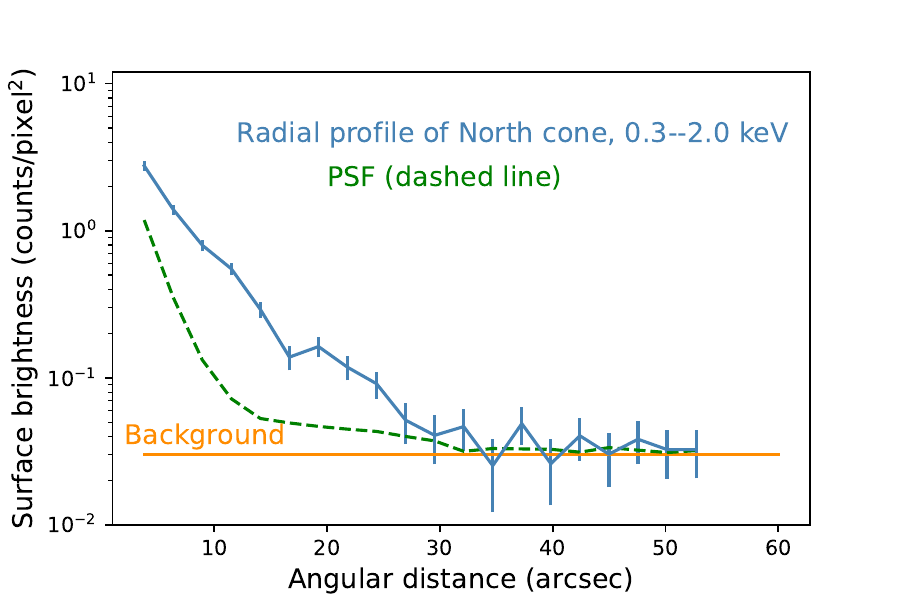}  
\includegraphics[width=10cm,clip]{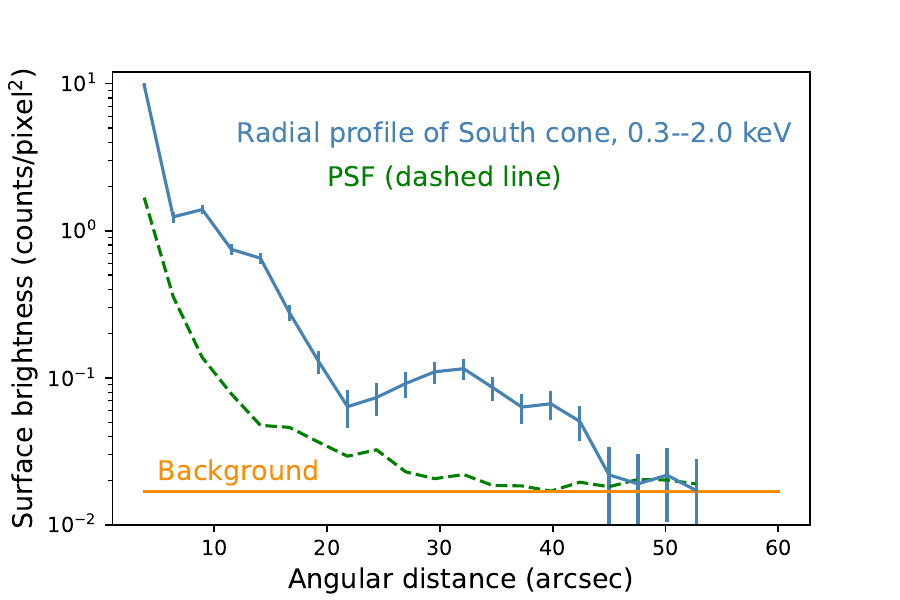}
}  
\caption{The radial profiles of the soft X-ray (0.3--2.0 keV) extensions in two directions separately. The upper and lower panels show the radial profile of the cone in the northwest and the cone in the southeast, respectively.}
\label{radialcone}  
\end{figure}

We further extracted the radial profiles of the 0.3--2.0 keV emission in the SE cone and NW cone separately.  Figure~\ref{radialcone} reveals that neither of the two radial profiles decrease smoothly. There are brightness bumps along the curves corresponding to the arcs apparent in Figure~\ref{cmap}a. The excess at $\sim 20\arcsec$ in the NW corresponds to the X-ray and optical nebular region hosting the ULX reported in \citet{2015ApJ...814....8K}. The radial profile of SE cone exhibits more structure than the NW, showing a gradual decrease in brightness with a cusp at  $r\sim 9\arcsec$ and a broad excess peaking at $r\sim 32\arcsec$ from the nucleus.

\begin{figure}[ht]
\begin{center}
\includegraphics[width=10cm,clip]{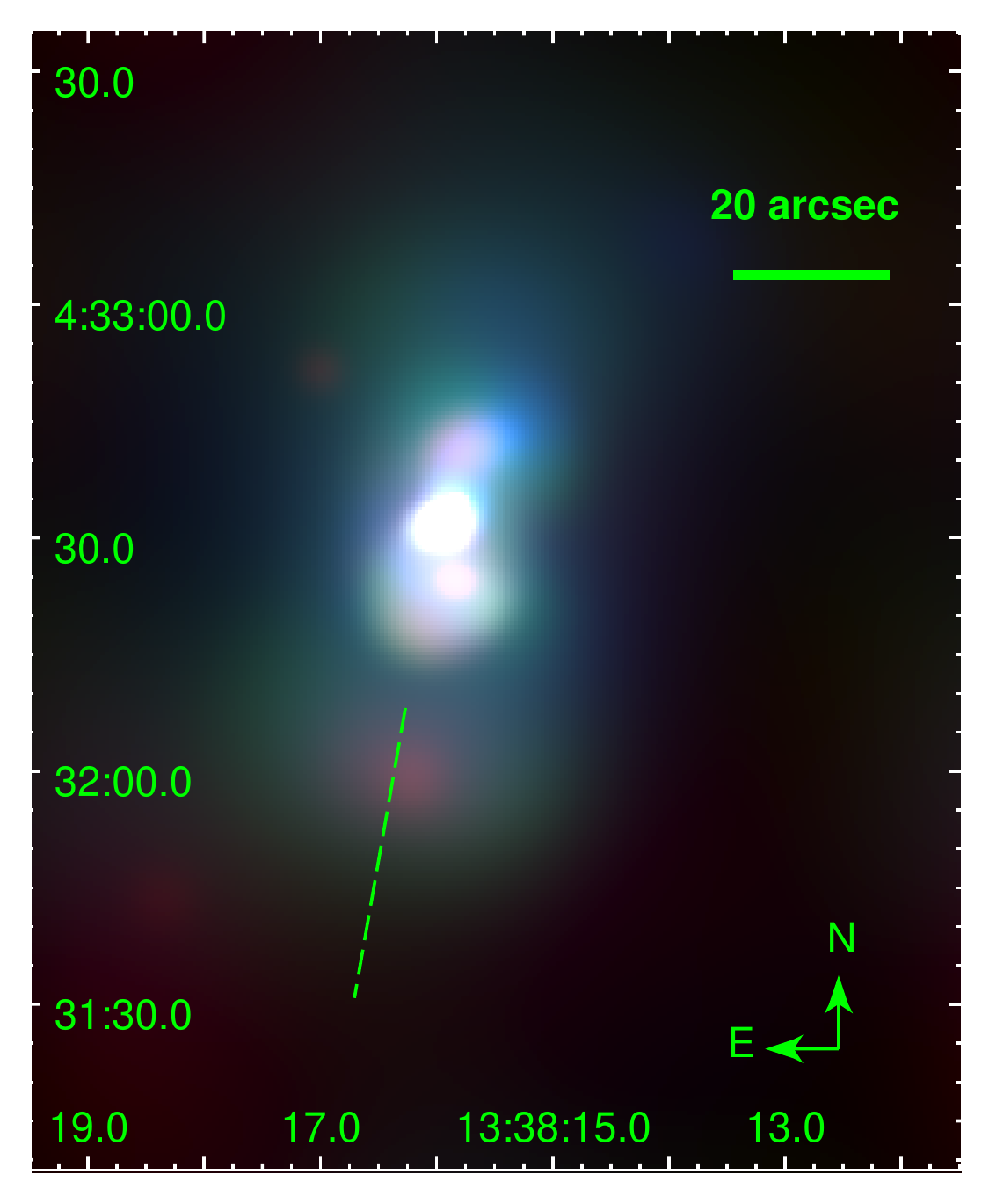}
\end{center}
\caption{Color composite image showing the morphology of the X-ray emission lines. Each color denotes one narrow energy band which covers one or more emission lines. Red corresponds to the 0.3--0.5 keV emission, green corresponds to the 0.5--0.6 keV emission, and blue the 0.6--0.8 keV emission. The dashed line shows the axis of the large scale soft X-ray extension. }
\label{nar_ext}
\end{figure}

Previous {\em XMM}-Newton study of \citet{2010A&A...516A...9D} reported the detection of X-ray emission lines including $O{VII}$ He-$\alpha$ triplet, $C{VI}$ Ly-$\alpha$, and $O{VIII}$ Ly-$\alpha$. To outline any structure dominated by these line emission, we extract three energy bands which are 0.3--0.5 keV, 0.5--0.6 keV and 0.6--0.8 keV, each of these energy band covering one of the aforementioned emission lines. The composite color image of the emission in three bands is shown in Figure \ref{nar_ext}.  Overall the extension from the nucleus along north-south direction is aligned with the large scale bi-cone. However, besides the bright nucleus, circumnuclear clumps could be seen in the map.  The appearance of a slightly elongated nucleus to the SE is due to a clump 3$\arcsec$ from the nucleus. Another relatively bright clump is located 8$\arcsec$ south of the nucleus.

\citet {1994AJ....107.1227W} identified radio emission spatially coincident to the optical emission line region, showing weak jet-like features extending $\sim$2$\arcsec$ ($\sim$900 pc) to north and south from the nucleus. This radio jet aligns well with the axis of the ionization cones, and could have an impact in the circumnuclear region and relate to the emission clumps.  We select an annulus centered on the nucleus with an inner radius of 2.5$\arcsec$ and an outer radius of 5$\arcsec$, and divided the annulus into two regions along PA$\sim$175 degree, which is perpendicular to the PA of the jet-like radio source. This two half-annuli hereafter will be denoted as S0 and N0 in the spectral analysis later (see Section~\ref{sec:spec}).

An adaptively smoothed image of the large scale soft X-ray emission is shown in Figure~\ref{soft_ext}, showing the spatial extent of the faintest feature.  Besides the circumnuclear regions N0 and S0, We further defined the concentric regions N1 and N2 in the northwest, S1, S2 and S3 in the southeast based on surface brightness (enclosed number of counts) for spectral analysis next (see Section~\ref{sec:spec}). 
    
\subsection{Comparison between the [OIII] Line Emission and the Soft X-ray Emission}

The soft X-ray emission shows several structure resembling arcs or shells. Previous {\em HST} images of NGC 5252 also show that the narrow emission line regions are spatially resolved and confined to several arc-like regions. \citet{1996ApJ...464..177A} analyzed the optical emission lines in the extended structures of NGC 5252 comprehensively, concluding that the extensions are ionized gas. Since the spatial scale of the extensions are larger than several tens of kpc, the ionization structure along the narrow line regions could vary at different projected distances from the nucleus, which deserves further investigation. 

We reduced archival {\em HST}/WFPC2 data to obtain two optical images of NGC 5252 taken in FR533N and F588N filters (propID 5426 and 5616, PI: Tsvetanov), following standard pipeline using the {\em astrodrizzle} software\footnote{\url{https://hst-docs.stsci.edu/drizzpac}}. In order to calculate the flux of the [OIII]$\lambda 5007$\AA\/ emission line covered by FR533N band, we first aligned the two images. The counts per second in the images were converted to flux in erg s$^{-1}$ cm$^{-2}$, corrected for the bandwidth of the filters. Then we subtracted the F588N image as continuum from the FR533N emission line image. The resolution of the image was subsequently degraded to yield similar resolution as {\em Chandra} ACIS-S. The continuum subtracted image demonstrates the morphology of the [OIII] emission region, shown in Figure~\ref{nar_OIII}. Several arc-like features are distinct, and each spatially corresponds to the soft X-ray extension we described in Section 3.2. 

The X-ray emission in the central 10$\arcsec$ region is shown in Figure~\ref{cir_nuc} with the contours of [OIII] line emission. A clumpy arc feature in the soft X-rays is more prominent towards the south, similar to the [OIII] emission.  In general, such X-ray and [OIII] spatial correspondence in the NLR has been previously reported in \citet{2006A&A...448..499B}, indicating a common origin in a photoionized medium.  To compare with previous work, we calculated the flux ratio of [OIII] and soft X-rays (0.3 keV--2.0 keV) in these regions covered by {\em HST} and {\em Chandra}. The resulting ratio of the soft X-ray flux to [OIII] emission flux in each regions are summarized in Table~\ref{ratio_OX}. 


\subsection{Spectral Analysis with Photoionization Modeling}\label{sec:spec}

\begin{figure}[ht]
\centering
\includegraphics[width=15cm,clip]{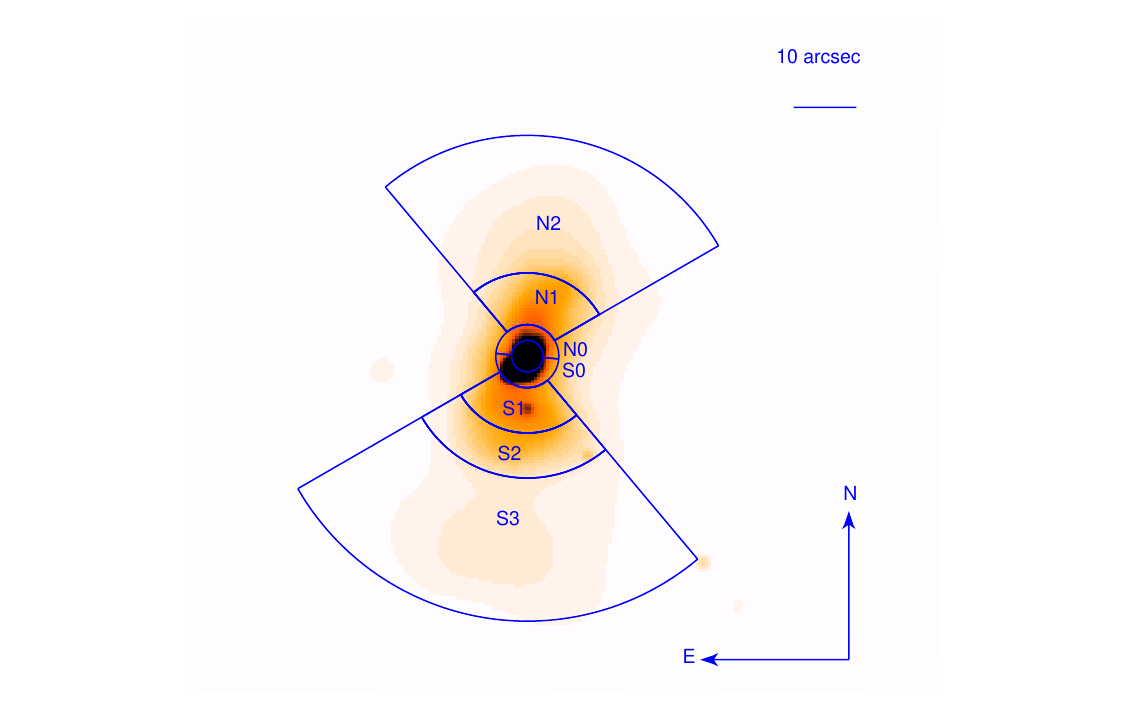}
\caption{Spectral extraction regions (sub-cones) of extended emission overlaid on the 0.3--2.0 keV image, which are defined based on the profiles exhibited in Figure~\ref{com} and Figure~\ref{radialcone}. The inner and outer radii of N0 and S0 are 2.5$^{\prime\prime}$ and 5.0$^{\prime\prime}$. Consequently the inner and outer radii of S1, S2, and S3 are 5$^{\prime\prime}$ and 12.2$^{\prime\prime}$, 12.2$^{\prime\prime}$ and 19.3 $^{\prime\prime}$, 19.3$^{\prime\prime}$ and 42$^{\prime\prime}$. For N1 and N2, the values are 5$^{\prime\prime}$ and 13.2$^{\prime\prime}$, 13.2$^{\prime\prime}$ and 35$^{\prime\prime}$. }
\label{soft_ext}
\end{figure}

\begin{figure}[ht]
\begin{center}
\includegraphics[width=10cm,clip]{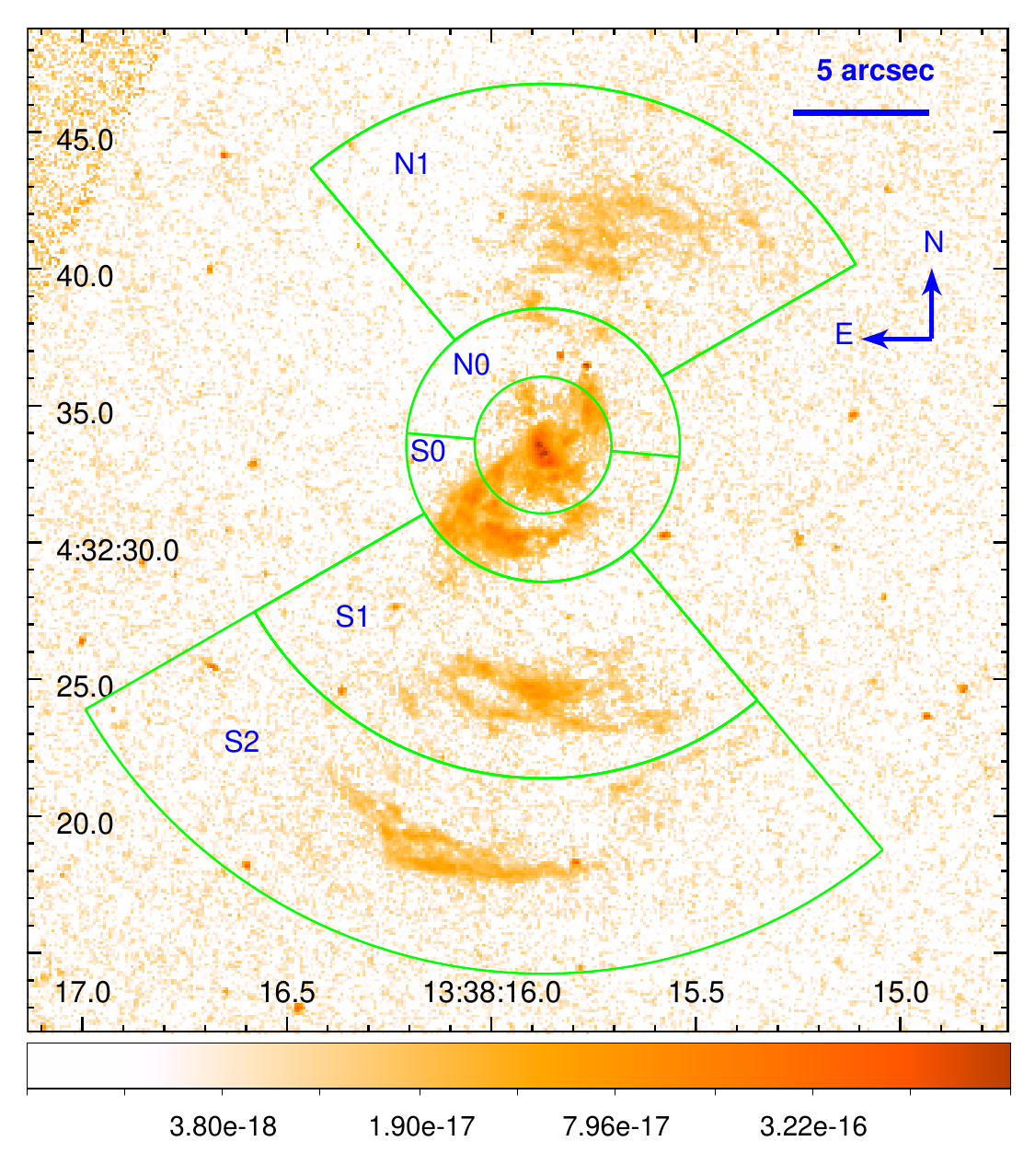}
\end{center}
\caption{Continuum subtracted {\em HST} image based on F533N and F588N, showing the morphology of [OIII] optical line emission. The X-ray sub-cone regions defined in Figure \ref{soft_ext} are spatially correlated with the [OIII] emission region.}
\label{nar_OIII}
\end{figure}

\begin{figure}[ht]
\centering
\includegraphics[width=12cm,clip]{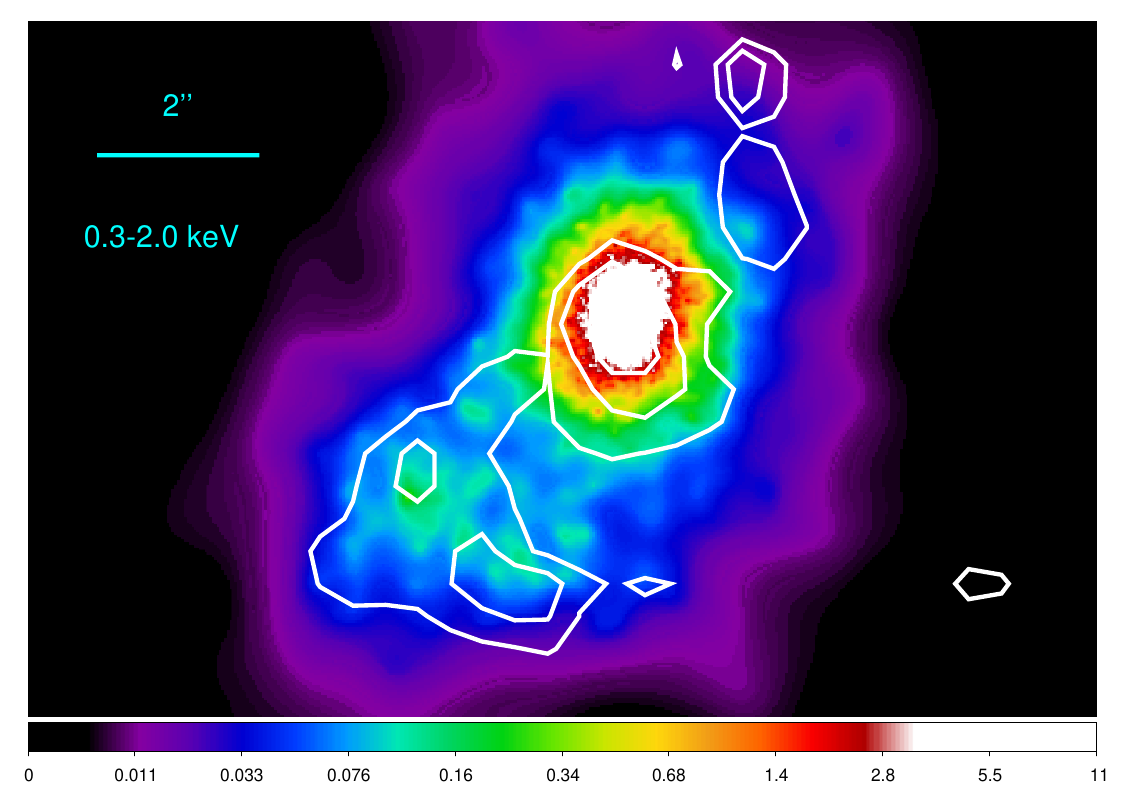}
\caption{{\em Chandra} ACIS-S image showing the nuclear region of NGC 5252 in soft X-ray band (0.3--2.0 keV). The image is overlaid with the contours of [OIII] emission from the continuum subtracted {\em HST} image.}
\label{cir_nuc}
\end{figure}

To perform spatially resolved spectral analysis of the extended X-ray emission, we first define the spectral extraction regions based on the structure of the cones.  Figure~\ref{soft_ext} shows that the extension in the 0.3--2.0 keV is up to 42$\arcsec$ from the center of the galaxy, corresponding to a physical scale of 19 kpc at the distance of NGC 5252.  
As described in Section~3.1, the radial profiles of bi-cones do not decrease smoothly and the SE cone reaches out further than the NW cone. An inflection appears in the angular distance of $16\arcsec$ in NW. The SE profile exhibits three features: it is relatively flat in the inner part ( $<13\arcsec$), becomes steeper from the angular distance of $\sim 14\arcsec$ to $\sim 22\arcsec$, and thereafter a bump arises at the large scale ( $\sim 22\arcsec$ to $\sim 42\arcsec$ ).

In order to take into account the radial profile structure described above, as well as ensuring sufficient number of counts in each region, the large-scale bicone was divided into five sub-regions, labeled as N0 -- N2 in the NW, and S0 -- S3 in the SE. Details of the partition information are shown in Figure \ref{soft_ext}.
 
The previous investigations based on optical \citep{1996ApJ...464..177A} and X-ray data \citep{2010A&A...516A...9D} of NGC 5252 conclude that the observed biconical extension surrounding the AGN is dominated by photoionized gas. The deep {\em Chandra} observations enable us to test this result with the X-ray spectra. Throughout the spectral fitting, the minimum absorption column density $N_H$ was fixed at $1.97 \times 10^{20}$ cm$^{-2}$, the value for the line-of-sight Galactic absorbing column towards NGC 5252 provided by the $CXC$ toolkit COLDEN\footnote{\url{https://cxc.harvard.edu/toolkit/colden.jsp}}. 
To explore the ionization properties of the soft X-ray extension, we generated models with the $Cloudy$ photoionization modeling code \citep[version C13.05;][]{2013RMxAA..49..137F,2017RMxAA..53..385F} to fit the spectra. We modeled the $<2$ keV X-ray spectrum assuming an open plane-parallel ``slab'' geometry.  photoionization model grids were computed and used in $XSPEC$. The ionization parameter $U$, and the column density of the slab $N_{\rm H}$, were varied to compute spectral models covering a grid of parameters.  
Here $U$ is the dimensionless ionization parameter \citep{2006agna.book.....O}, defined as $U = Q/(4\pi r^2 c n_{\rm H})$, where $Q$ is the rate of hydrogen ionizing photons ($s^{-1}$) emitted by the AGN, $n_{\rm H}$ the hydrogen number density, $r$ the distance to the inner face of a model slab, and $c$ the speed of light. For the AGN continuum, four parameters were specified, the temperature of AGN ``Big Bump'' component $T_{BB}$ = $10^6 $ K, the optical to X-ray spectral index $\alpha_{ox}$= $-$1.3, the low-energy slope of the Big Bump continuum $\alpha_{uv}$=$-$0.8, and the slope of the X-ray continuum $\alpha_x$=$-$0.8 \citep{1997ApJS..108..401K}. We simply assume the slabs of clouds in different regions are illuminated by exactly the same nuclear continuum. The abundance is set at the default solar composition used by Cloudy \citep{2013RMxAA..49..137F}. Notably, although our spectral analysis is based on a specific ionizing SED as described above, we have evaluated the impact of different AGN SEDs on our fitting results.  After extensively fitting with model grids generated using a range of $\alpha_x$ and $\alpha_{uv}$, we confirm that different SEDs yield comparable fitting results such that our results on the variation of ionization parameters are robust.
              
\begin{figure}[ht]  
\centering  
{  
\includegraphics[angle=270,width=8cm,clip]{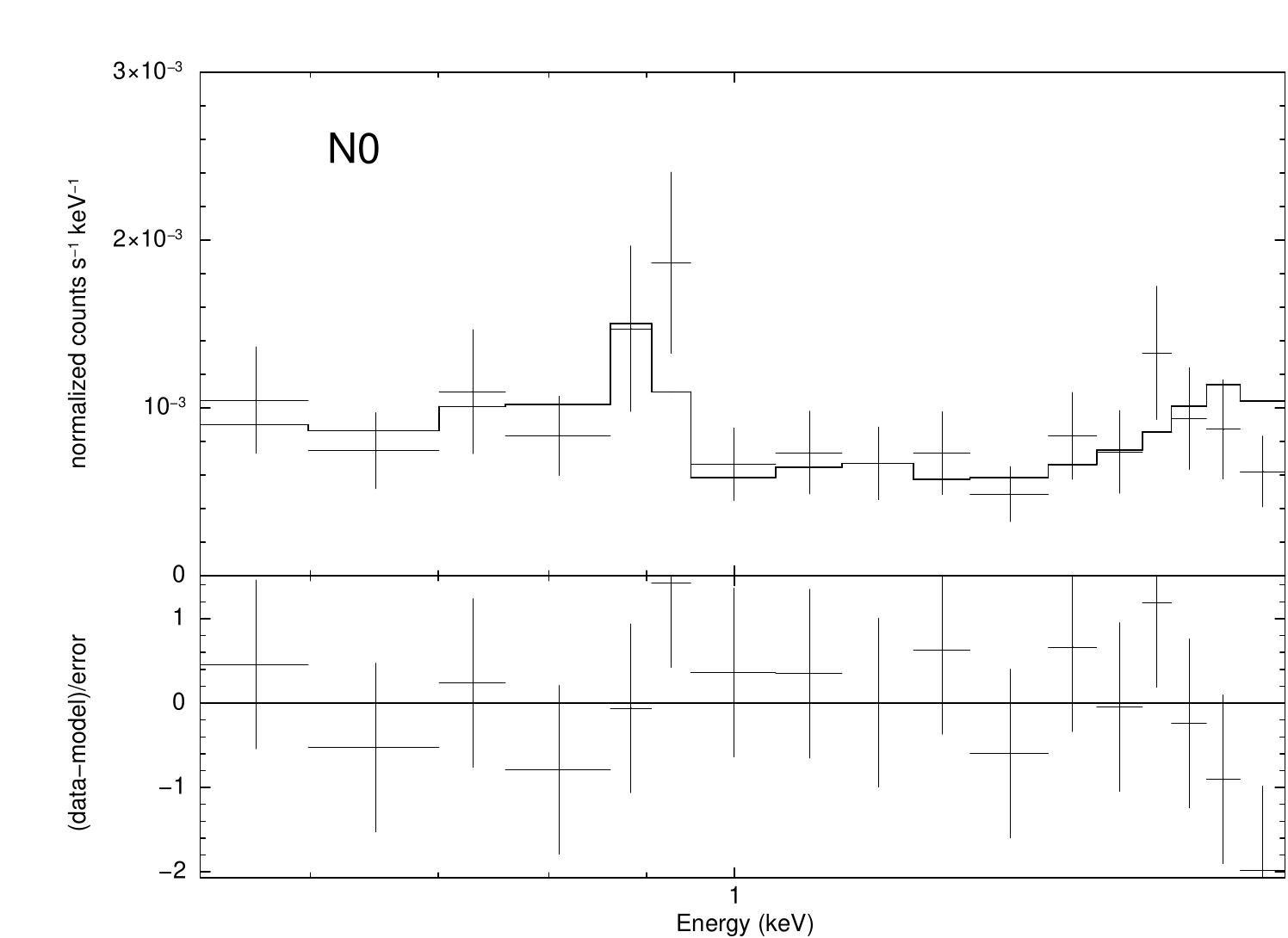}
\includegraphics[angle=270,width=8cm,clip]{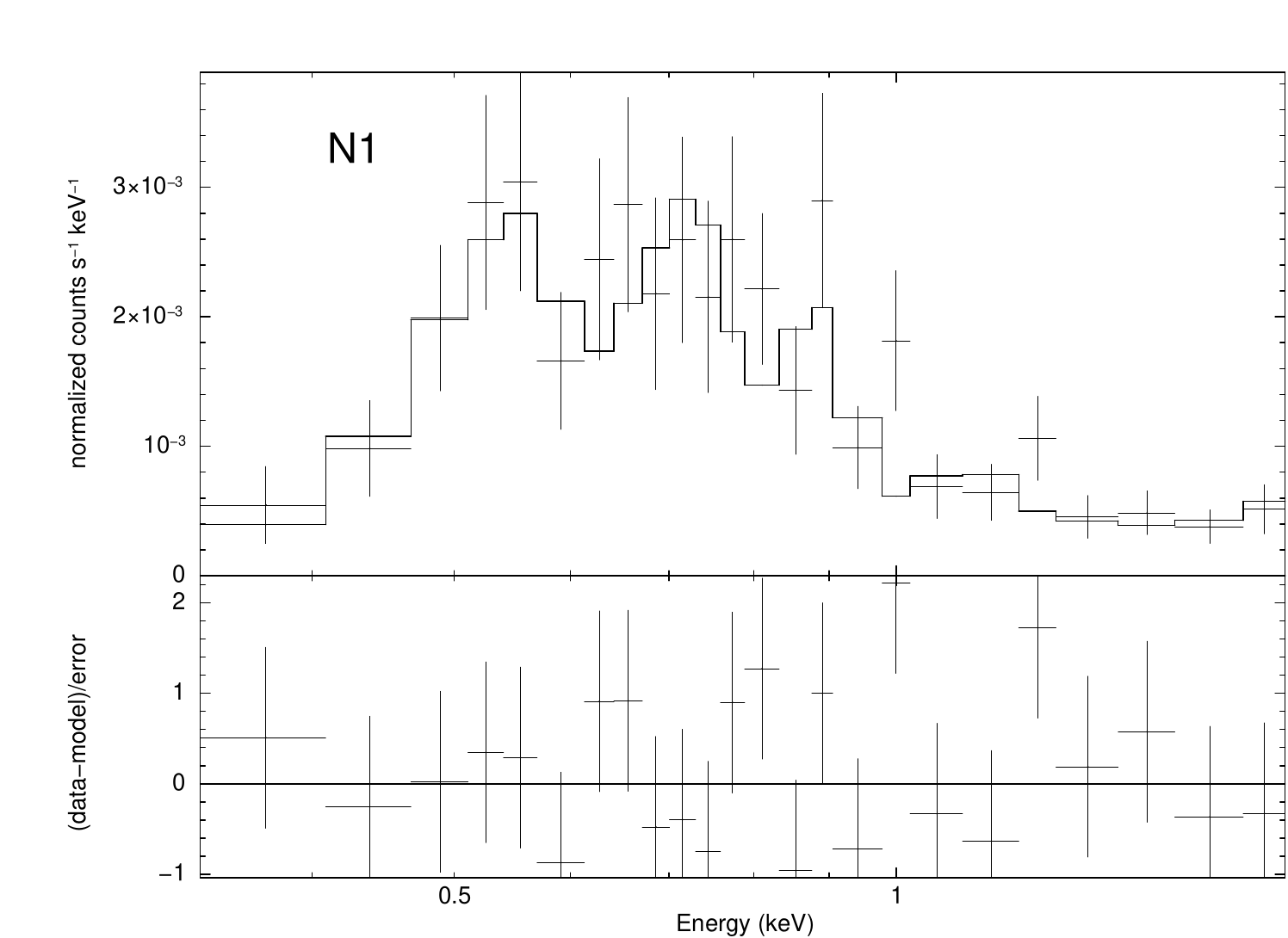}
\includegraphics[angle=270,width=8cm,clip]{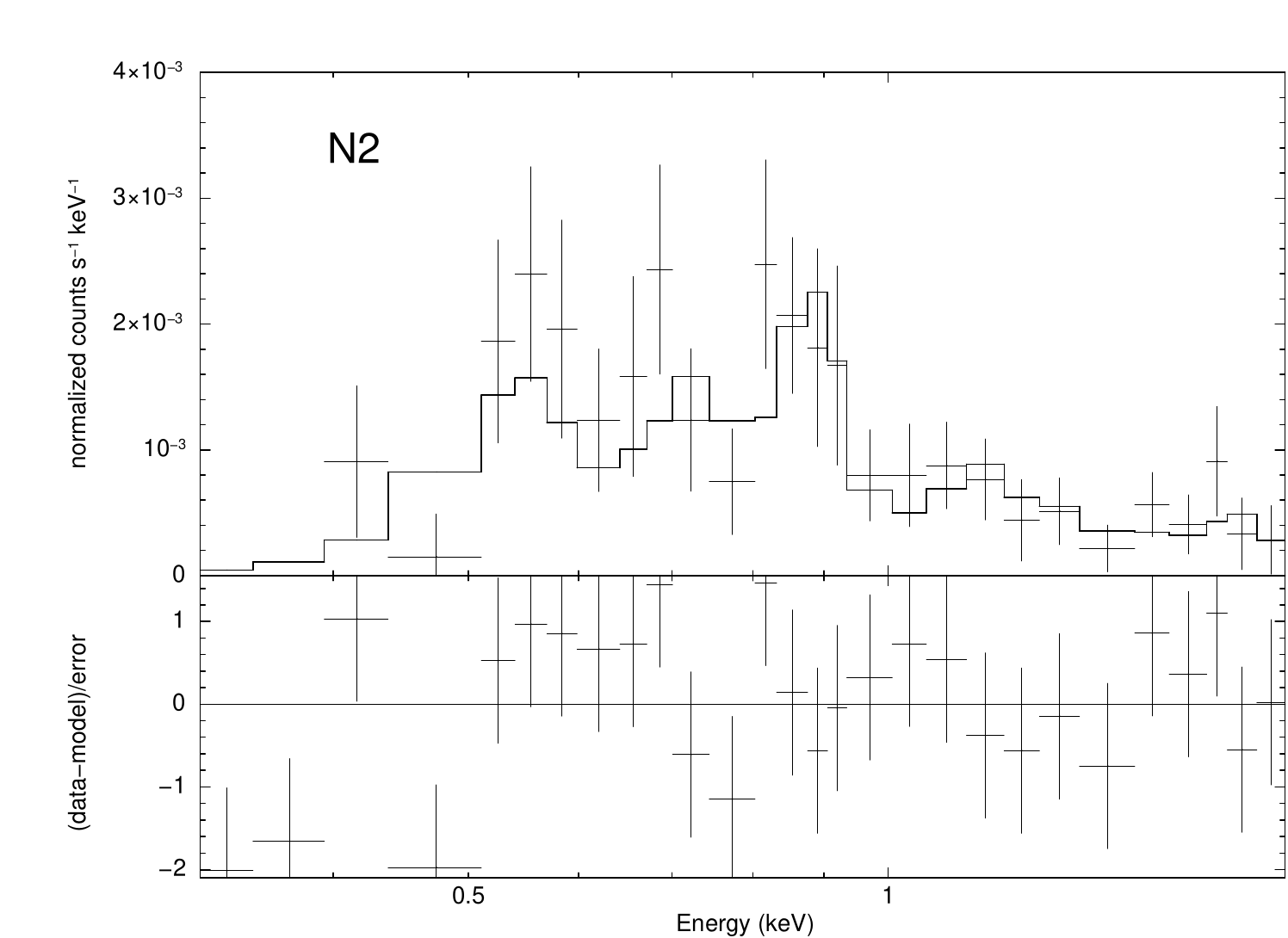}
}  
\caption{The spectra of the northeast sub-cone regions are modeled with the absorbed photo ionization model. The spectra extraction regions for the 3 spectra are N0 (left) and N1 (right) in the upper panel, and N2 in the lower panel.}  
\label{spectra_ne} 
\end{figure}

\begin{figure}[ht]  
\centering  
{  
\includegraphics[angle=270,width=8cm,clip]{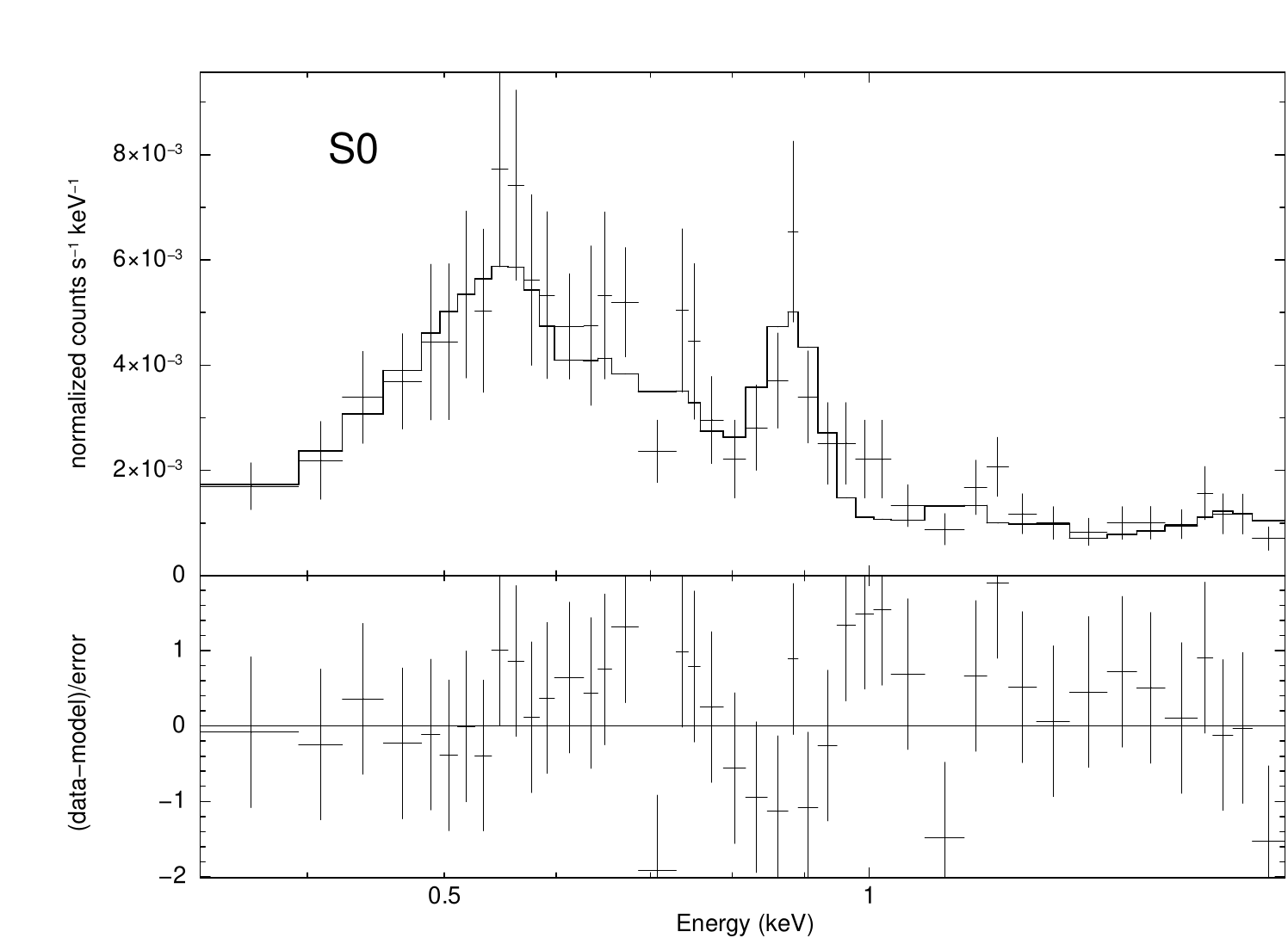}  
\includegraphics[angle=270,width=8cm,clip]{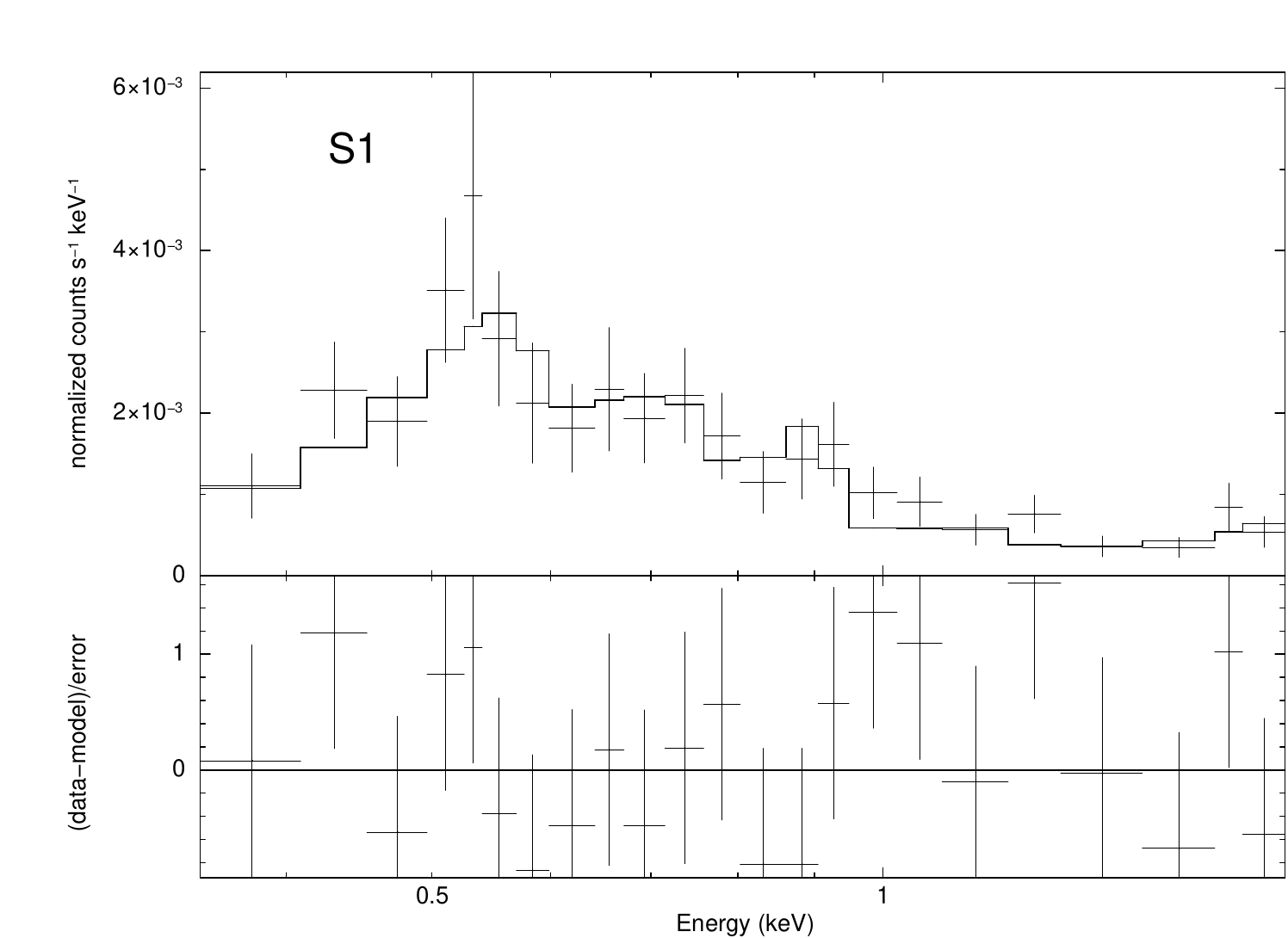} 
\includegraphics[angle=270,width=8cm,clip]{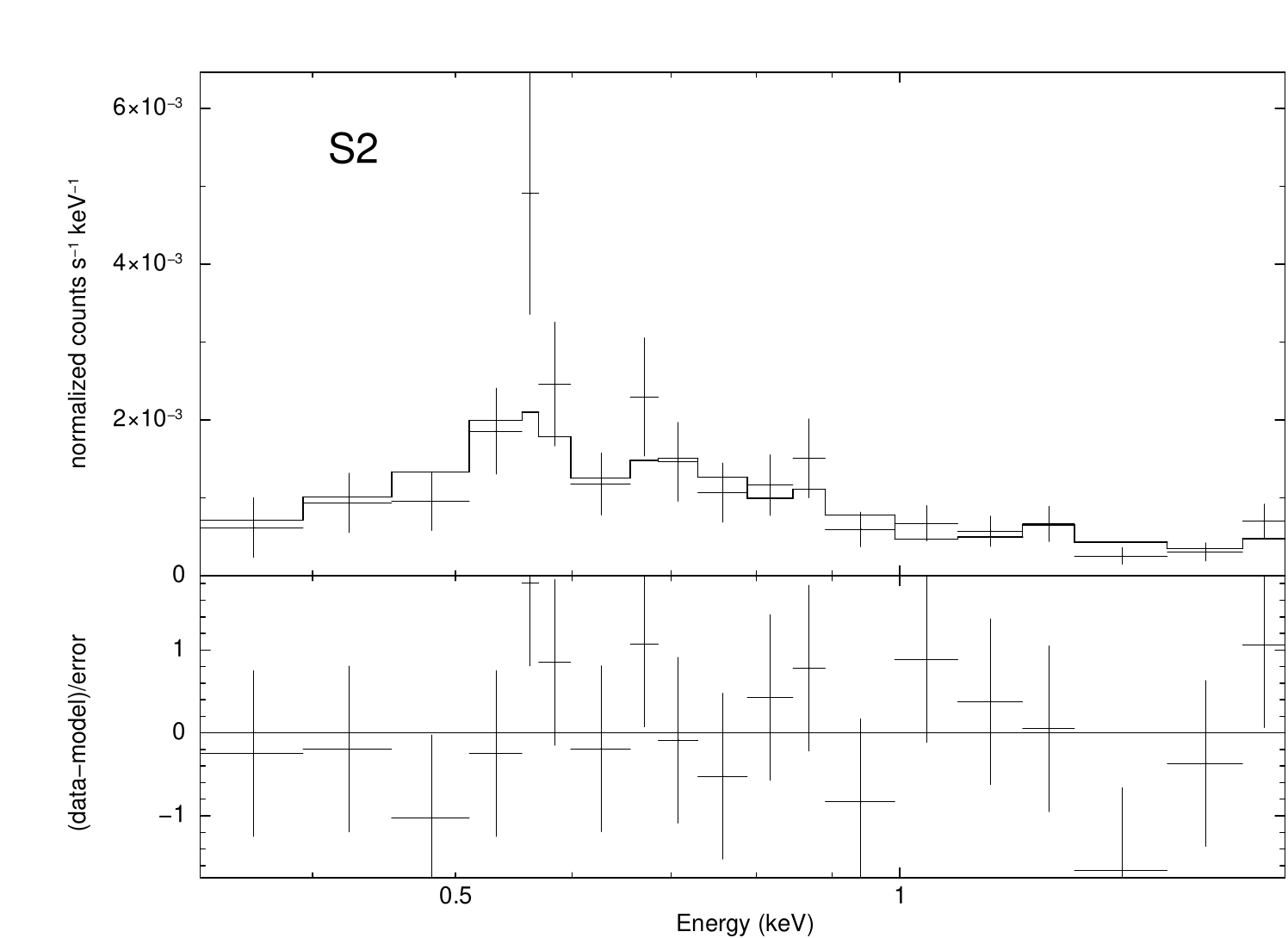}
\includegraphics[angle=270,width=8cm,clip]{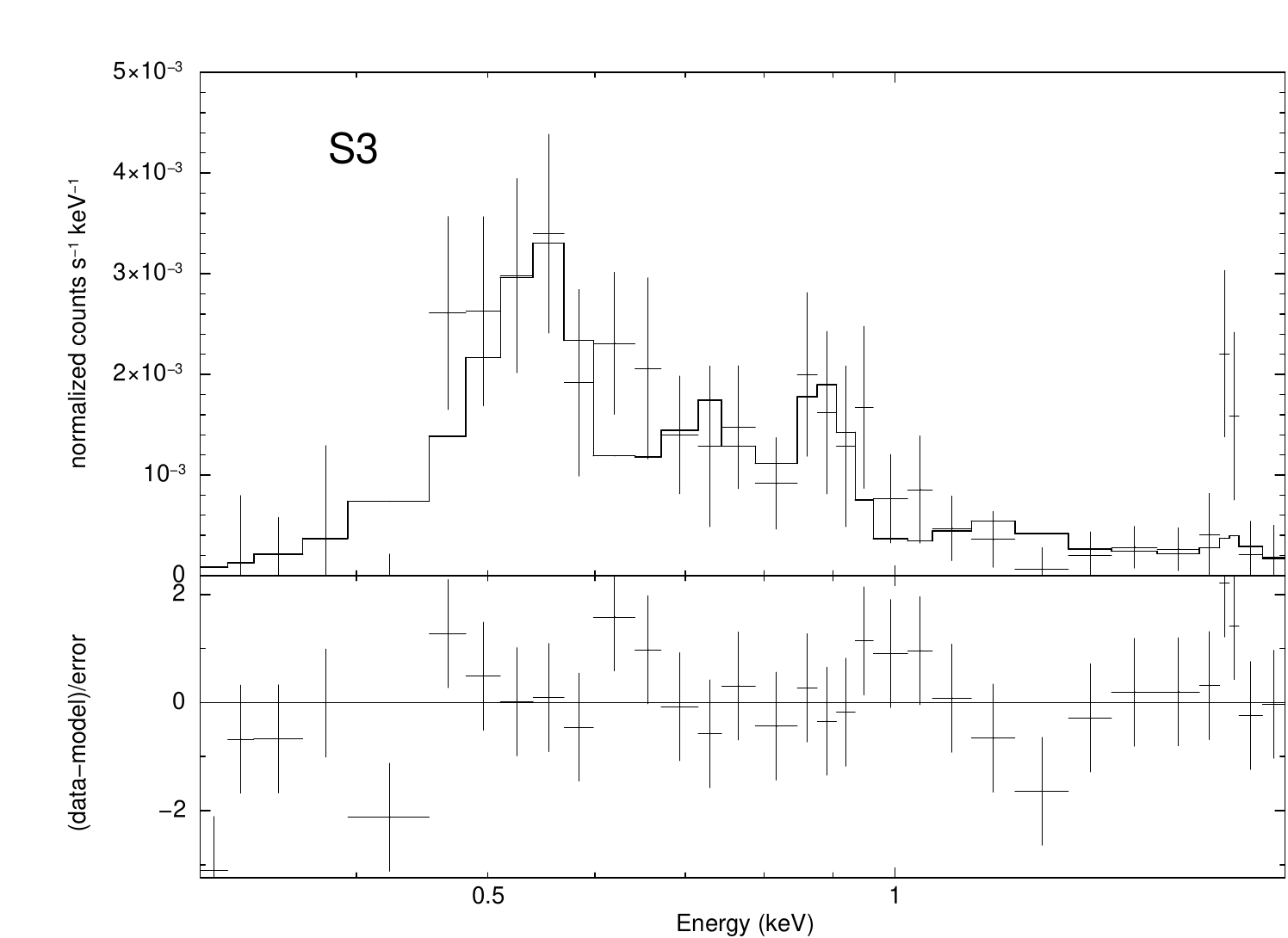}
}  
\caption{The spectra of the southwest sub-cone regions are modeled with the absorbed photo ionization model. The spectra extraction regions for the 4 spectra are S0 (left) and S1 (right) in the upper panel, S2 (left) and S3 (right) in the lower panel.}  
\label{spectra_sw}  
\end{figure}

It is expected that the spectra in regions close to the nucleus (namely N0, S0, N1, and S1) are subjected to notable contamination from the scattered AGN continuum in the PSF wings. This was properly taken into account in the spectral fitting by including the nuclear spectrum as a background component, scaled by the corresponding PSF fraction calculated from the simulated PSF image. The spectra can be well fitted by this photoionization model, except that S2 region requires a second photoionized component with a lower ionization parameter which significantly improves the fit ($\Delta \chi^{2} = 5.42$). It could be ascribed to the possible existence of local denser gas in this area. Although S2 region does not apparently show more clumpy structure in the optical image, this assumption is supported by evidence from previous optical studies. \citet{1998ApJ...505..159M} identified a unique narrow filament labelled as SH15 residing in the south of the nucleus of NGC 5252, which spatially corresponds to S2 in our work. The ionization parameter of SH15 is about two times lower than the southeast spiral arm. This result is also supported by the decrease of [O III]/H$_{\alpha}$ ratio, which is also an indicator of ionization level, in SH15 ($\sim$3.3) compared to the southeast spiral arm ($\sim$4.5). Moreover, the analysis of the emission lines by \citet{1996ApJ...464..177A} reveals strong high-ionization lines as well as low-ionization lines in SH15 which indicates the presence of a low ionization component in this region. The fitting parameters are summarized in Table~\ref{ionization}, and the fluxes listed in Table~\ref{ionization} are referring to the photoionized components. The fitted spectra are shown in Figures~\ref{spectra_ne} and~\ref{spectra_sw}.
The modeling results in Table~\ref{ionization} indicate that the absorption columns $N_H$ in the northwest cone are about one order of magnitude higher than that of the southeast region, which implies higher obscuration to the north. This is likely caused by the inclination of the galactic disk, the north side of the nucleus would suffer more obscuration from the stellar and gas disk of the galaxy. Although with some variations, the ionization parameters log$U$ are comparable across the regions considering the uncertainties. 

\begin{figure}[ht]
\centering  
\includegraphics[width=12cm,clip]{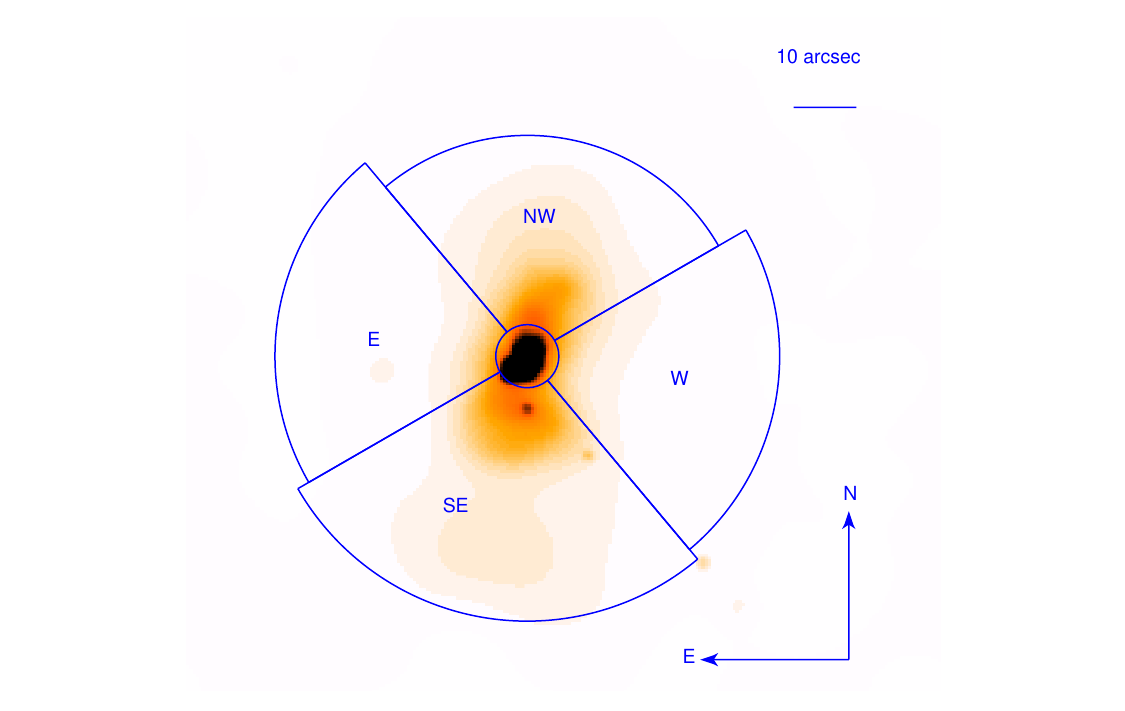}
\caption{The regions overlaid on the 0.3--2 keV soft X-ray image defining the four cones (NW, SE, E, and W) which are described in Section \ref{sec:large_cone}. Regions E and W correspond to the lower ionization cones.}
\label{sidecone}
\end{figure}

For comparison, the spectra from two regions in the vicinity of the bi-cone were extracted, denoted as E and W respectively in Figure \ref{sidecone}. These two side cones are spatially complementary with the soft X-ray extensions. Then we fitted the spectra with a photoionization model and compare the fitting results of the side cones (E and W) and the sub-cones of the extensions (N1--N2, S1--S3). Notably, each cone contains $ \sim 0.4\%$ nuclear scattering photons based on the simulated PSF, therefore we added the normalized nuclear component to the model. The fitting parameters of the two spectra are listed in Table \ref{wholeph}.




\section{Discussion}

\subsection{The Origin of Extended Soft X-rays}
\label{sec:large_cone}

Owing to the excellent spectral resolution of RGS aboard $XMM$-Newton, \citet{2010A&A...516A...9D} analyzed the intensity of several emission lines from the vicinity of the NGC 5252 nucleus and the bi-cones. The line diagnostic favors the photoionization mechanism rather than collisional ionization. A similar case is IC 5063, for which the study of \citet{2021ApJ...921..129T} and \citet{2023MNRAS.520.1848H} shows evidence of AGN photoionization in kpc spatial scale.  

 Following the conclusions of previous studies, the spectral modeling analysis in Section~\ref{sec:spec} assumes that the observed soft X-ray extensions of NGC 5252 are dominated by photoionization. Alternatively, collisionally ionized hot gas heated by shocks driven by AGN outflow is another competitive scenario for the extended soft X-ray emission in NGC 5252, and needs to be investigated. There is observational evidence showing that outflows or jets from AGN could interact with the host ISM reaching tens of kpc scale. For example, in the radio quiet quasar known as the ``Teacup'' AGN \citep{2015ApJ...800...45H}, {\em Chandra} imaging data resolves a 10 kpc loop structure with properties of X-ray emission in good agreement with a shocked thermal gas \citep{2018ApJ...856L...1L}. At smaller spatial scale (tens or hundreds of pc), radiation mechanism of extended soft X-rays could be more complicated, it could be the mixture of collisional ionization and photoionization as found (see Fabbiano \& Elvis 2022 for a review), for example, in NGC 1068 and NGC 4151 \citep{2023MNRAS.524..886H, 2011ApJ...742...23W}.

We further explore the possible presence of hot gas using spectral analysis. The spectra were extracted from two large cones, SE (combination of S1, S2, and S3) and NW (combination of N1 and N2) respectively, ensuring each cone covers the whole extension (see Figure~\ref{sidecone}). The spectra are fitted with an absorbed thermal plasma model $apec$. The PSF scattering was again taken into consideration, with $0.4\% $ of the nucleus contaminating each region. The thermal model also provides statistically acceptable fits ($\chi^2/d.o.f$ is $44.47/57$ in the SE and $54.73/47$ in the NW). Moreover, the fitting parameters show the temperatures of the two regions are in reasonable range. The best-fit $kT$ in NW is  0.15$\pm$0.01 keV. The spectrum of SE requires two thermal components, and the temperatures are 0.54$\pm$0.17 keV and 0.08$\pm$0.02 keV. The presence of collisionally ionized hot gas cannot be ruled out solely based on the spectral fitting.  Considering the limited spectral resolution of {\em Chandra} ACIS, it is not totally surprising the spectral fitting cannot distinguish between the photoionization and the collisional ionization models. 


\citet{1989Natur.341..422T} claimed that there is a low ionization zone perpendicular to the ionized extension, across the nucleus of NGC~5252 (see Table~\ref{oriention} for the PAs of various features). It firmly supports that the nucleus radiation is anisotropic and the collimated radiation field is consistent with the direction of the soft X-ray extension. Therefore, it is straightforward to assume that, in case of the anisotropic photoionization scenario, the ambient region of the bi-cone would show lower ionization due to a heavily blocked ionization continuum. This hypothesis could also be verified by the spectral analysis. The spectral fitting shows that the two side cones E and W are indeed characterized by lower ionization parameters compared to the X-ray ionization cones. Moreover, the column density $N_H$ are also very low ($\sim10^{19}$ cm$^{-2}$). The results are consistent with the scenario that the ionizing radiation did not reach farther because of the existence of obscuration, possibly from the torus \cite{1989Natur.341..422T} surrounding the nucleus. Hence only a thin layer of gas close to the nucleus in region E and W is ionized, with lower ionization parameters.

Besides the X-ray properties, the dynamics of the gas disk encompassing the galaxy revealed by optical and radio observation preferentially supports the photoionization scenario. \citet{1996MNRAS.279...63P} argued that the velocity fields of the ionized gas and the ambient neutral gas are similar and it is likely that the two components belong to one complex rotation plane. If the ionized gas is part of the hot outflow originated from the nucleus, it is unlikely to have the same kinematics with the ambient neutral gas. Moreover, \citet{1998ApJ...505..159M} claimed the existence of three large gas disks surrounding the galaxy. The gas disks are partially illuminated by the anisotropic nuclear radiation of NGC 5252. This possibly resulted in the observed morphology of [O III] arcs and filaments, as part of the shell or debris from past mergers is irradiated.  In short, all the morphological, spectral and kinematic information are consistent with a photoionized origin for the extended soft X-rays, whereas the hot outflow scenario is not a favored explanation. 


\subsection{The Luminosity Variation History of the NGC 5252 Nucleus}

\begin{figure}[ht]
\centering
\includegraphics[angle=0,width=12cm,clip]{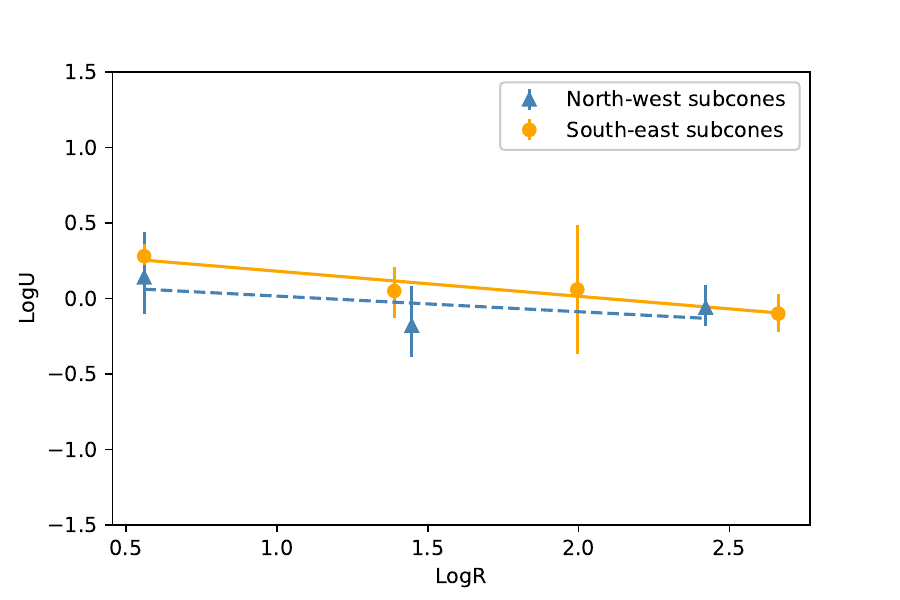}
\caption{Comparison between measurements and the linear model fit to the relation of log$U$ and log$R$ in all the sub-cone regions. log$U$ is the ionization parameter in photoionization model. $R$ is the radii from the outer layer of each sub-cone to the center of NGC 5252 (in unit of arcsecond). The filled circles denote the data of the southwest sub-cones (S0, S1, S2, and S3) and the triangles denote the northeast sub-cones (N0, N1, and N2).}
\label{tendencylogu}
\end{figure}

 The flux ratio of [OIII] to soft X-ray is an indicator of the ionization degree in narrow line region. \citet{2006A&A...448..499B} and \citet{2009ApJ...704.1195W} summarized the [OIII] to soft X-ray ratio of the narrow line regions of several Seyfert galaxies or galaxies in pair. The ratios are mainly between 1 to 10 \citep{2006A&A...448..499B,2019ApJ...884..163F}.  
 The flux ratio of [OIII] to soft X-ray measured in NGC 5252 are $\sim$ 2--7 (Table~\ref{ratio_OX}), which are consistent with the values in NLR of several Seyfert Galaxies. Notably, in the case of NGC 5252, the flux ratios along the extensions are quite stable considering that the NLR regions already reach out to several kpc.
Figure~\ref{tendencylogu} shows the radial variations of the log$U$ with respect to log$R$, where $R$ is the projected distance from the central position of each sub-cone region to the nucleus. The results manifest that, despite the large extension of the photoionized gas (reaching out to about 19 kpc), its ionization parameter does not significantly decrease as the radius increase. The data log$U$ and log$R$ were fitted with a linear function (U $\propto$ $R^{\alpha}$), the fitting results are U $\propto$ $R^{-0.2 \pm 0.1}$ in southwest and U $\propto$ $R^{-0.1 \pm 0.2}$ in northeast (R is in the unit of kpc).  The best-fit power indices ($-0.2$ and $-0.1$) are quite small, implying that the ionization degrees in each sub-cone are comparable.  We have extensively verified that adopting different AGN SEDs to generate the photoionization model grids yields consistent values of $\alpha$ within uncertainties. This is further supported with the rather constant ratios between the [OIII] and soft X-ray fluxes.




Adopting the relation U = $\frac{L}{4 \pi \rho R^{2} }$ here, we could estimate how the luminosity of the nucleus may have varied during the past 64000 years (the light travel time over the $\sim$20 kpc scale). Since the luminosity is also sensitive to the density $\rho$ of the ionized gas, the key step is to figure out the distribution of the number density along the bi-cone. 

Fortunately, the optical data offered some clue of the physical property of the ionized gas. The previous work of \citet{1998ApJ...505..159M} and \citet{1996ApJ...464..177A} both show that the electron density in different emission line regions are basically constant. If assuming that the X-ray bicone share the same origin with the emission lines regions, $\rho \propto R^{0} $ would be a reasonable description. 


We can use the radial dependence of the ionization parameter along the cone to estimate the temporal evolution of the AGN X-ray luminosity, under the assumption that the structures in the X-ray ionization cone are primarily driven by the instantaneous ionizing flux. Assuming that $U \propto R^{-0.2}$ (as in the South cone), the definition of $U$ yields $L \propto \tau^{1.8}$, where $\tau \equiv R/c$. We consider here the luminosity at the outermost regions of the South cone, which provide the longest time baseline and the most accurate determination of the ionization parameter due to lower obscuration.  The difference in ionizing luminosity between S0 (at 2.3~kpc) and S3 (at 19~kpc) is a factor of 46 $\pm$ 10. The light crossing times correspond to 7,600~years and 64000~years, respectively.  A historical large variation of the intrinsic AGN emission is also consistent with the previous work which finds that NGC 5252 went through a galaxy merger and it used to be a quasar instead of a Seyfert 2 \citep{2005A&A...431..465C}. Some theoretical simulations expect that AGNs experience episodes of such a state transition accompanied with highly different luminosities when the accretion rate changes \citep[e.g.,][]{2011ApJ...737...26N,2019ApJ...885...16Y,2023MNRAS.523.1641D}.
 
\begin{figure}[ht]
\centering  
\includegraphics[width=12cm,clip]{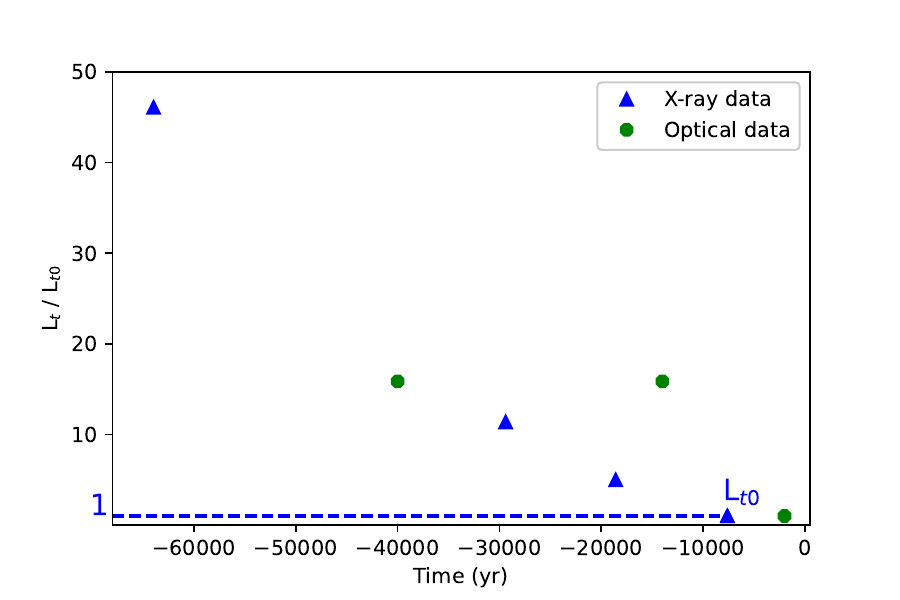}
\caption{The inferred variation history of the luminosity of AGN over the time span of $\sim$ 60000 yr. $L_{t}$ denotes the luminosity of AGN in each epoch, $L_{t0}$ is its luminosity in $\sim$ $-$7,600 yr. The vertical axis is the ratio of $L_{t}$ comparing with $L_{t0}$. The blue triangles shows the variation of nuclear luminosity during $-$ 64000 to $-$7,600 yr based on X-ray data. The green filled circles are the results from \cite{2017ApJ...835..256K} based on optical data, which covers the time span from $-$40000 to $-$2000. }
\label{luminosity}
\end{figure}

This result can be compared with similar estimates derived from spatially-resolved spectroscopy of the ionization cone in the optical. The study of \citet{2017ApJ...835..256K} presented the inferred long term behavior of the nuclear luminosity of NGC 5252. It had experienced three phases. The nuclear luminosity remained stable during $\sim -$40000 yr to $-$14000 yr. Then it began to decrease from $\sim-$14000 yr to $-$ 2000 yr with an index of 1.2, where the index is defined as $\log(\frac{L_{t_{0}}}{L_{t}}$). This index quantifies the variation of the luminosity after a certain duration of time. $L_{t_{0}}$ is the luminosity at the starting epoch corresponding to inner-most measurement, and ${L_{t}}$ is the luminosity in the past. And during the past 2000 years, it faded with the index of 1.4. In total, the nucleus used to be 46 times brighter than its current observed luminosity. Our data, which covers the time span $\sim -$64000 yr to $-$7,600 yr, roughly correspond to the first two phases ($\sim -$40000 yr to $-$2000 yr) and yield comparable results. Figure \ref{luminosity} shows the inferred variation of nuclear luminosity based on our modeling results and the comparison with the results of \cite{2017ApJ...835..256K} based on optical measurements. Some difference between the slopes of variations could be attributed to the crude assumption of constant density profile for X-ray gas, as they are in different ionization state from the optical ionized gas and may not follow the same exact spatial variation.
  
\section{Conclusion}

In this work, we analyzed the X-ray properties of NGC 5252 using {\em Chandra} ACIS-S data. The spatial extent, the morphology, and the dominated energy band of the X-ray emission are studied. We carried out a spatially resolved spectral analysis of the large X-ray extensions of NGC 5252. The main findings are summarized as follows.

Owing to the depth of our X-ray imaging data, it is the first time that the X-ray extensions reaching out to 19 kpc from the nucleus of the NGC 5252 galaxy were detected. Thanks to the excellent spatial resolution of {\em Chandra}, the X-ray cone structures are spatially resolved, mostly follow the direction of the optical ionization cones in the SE and NW direction. The dominated energy band of the extended emission is 0.3--2.0 keV. Lower ionization gas are also found in the E and W side of the cones. The spectral analysis with {\em Chandra} data and lacking of kinematical evidence for outflow supports the previous conclusion that the extended soft X-rays are due to photoionization and the ionizing source is the active nucleus of NGC 5252.

As the radial distance of the extended structure from the nucleus increases, the photoionization parameter of ionized gas in the sub-cones stays constant or slightly decreases. Due to light travel time, it takes more time for the ionizing continuum from nucleus to reach farther. Therefore the photoionization degree in each sub-cone reflects the luminosity of the nucleus in different epoch. Assuming that the density along the extension is uniform on the 10 kpc galactic scale, the X-ray measurement shows that the inferred luminosity of the nucleus used to be higher than its current luminosity. More quantitatively, it may have decreased about 98\% during the past 64000 years. A possible scenario for the observational phenomenon in NGC 5252 is that the galaxy went through a minor merger event which significantly influenced the nucleus in terms of its luminosity. The scenario of hot outflow from the central nucleus for the extended emission is much less favored. 

Together with IC 2497 and other ``Hanny's Voorwerp'' objects \citep{2022ApJ...936...88F}, NGC 5252 offers a great opportunity to understand the transition of states between high accretion rate and low accretion rate for AGNs. The significant variation on the timescale of $10^4-10^5$ years is consistent with predictions of both numerical simulations \citep{2019ApJ...885...16Y} and observational constraints \citep{2015MNRAS.451.2517S}, in which episodes of high accretion occur due to elevated gas supply during the growth of SMBHs. This implies many cycles of such accretion outbursts can be expected while the black holes and the host galaxies co-evolve, which leaves observable imprint of AGN feedback on the gas in the galaxy and on circum-galactic scale. Similar to discovery of the large scale {eROSITA} bubbles in our Milky Way \citep{2020Natur.588..227P}, more detections in local galaxies are envisioned and provide further insight in the past activities of their SMBHs and suppression of star formation.


\section{Acknowledgments}

We sincerely thank the anonymous referee for many helpful suggestions. J.W. acknowledges the National Key R\&D Program of China (Grant No. 2023YFA1607904). We acknowledges the NSFC grants 12033004, 12221003, 12333002 and the science research grants from CMS-CSST-2021-A06. C.W. acknowledges support by CNES under an XMM/SSC contract. This research has made use of data obtained from the {\em Chandra} Data Archive, and software provided by the CXC in the application packages CIAO and Sherpa. This research used observations made with the NASA/ESA {\em Hubble Space Telescope}, and obtained from the Hubble Legacy Archive, which is a collaboration between the Space Telescope Science Institute (STScI/NASA), the Space Telescope European Coordinating Facility (ST-ECF/ESA) and the Canadian Astronomy Data Centre (CADC/NRC/CSA).  This paper employs a list of {\em Chandra} datasets, obtained by the Chandra X-ray Observatory, contained in~\dataset[DOI: 10.25574/cdc.181]{https://doi.org/10.25574/cdc.181}.
 {\em HST} data presented in this paper were obtained from the Mikulski Archive for Space Telescopes (MAST) at the Space Telescope Science Institute. The specific observations analyzed can be accessed via \dataset[DOI:10.17909/9y0r-9n35]{ http://dx.doi.org/10.17909/9y0r-9n35}.

\facilities{CXO (ACIS), HST (WFPC2)}


\software{astropy \citep{2013A&A...558A..33A,2018AJ....156..123A,2022ApJ...935..167A}, CIAO \citep{2006SPIE.6270E..1VF}, DS9 \citep{2003ASPC..295..489J}
          }

\clearpage

\clearpage

\clearpage

 \clearpage

 \clearpage

\clearpage

\clearpage

\clearpage

\clearpage

\clearpage

 \clearpage

 \clearpage

 \clearpage

 \clearpage

\begin{table} 
\caption{Log of {\em Chandra} Observations of NGC 5252}  
\begin{center}  
\begin{tabular}{ccccc}  
\hline \hline    
ObsID & PI & Date  & Nominal Roll Angle [deg] & Duration [ks]\\ 
\hline  
4054 & Dadina &August 11, 2003  & 255.2 &59.8\\ 
\hline  
15618& Wang & March 4, 2013 & 80.2 &40.4 \\ 
\hline 
15022& Wang & March 7, 2013   & 82.2 &67.6  \\
\hline 
15621& Wang & May 9, 2013  & 223.8 &62.3  \\
\hline \hline  
\end{tabular} 
\end{center}
\label{io}
\end{table}

 \clearpage

\begin{table} 
\caption{Flux of [OIII] and soft x-ray in each sub ionized gas zone}  
\begin{center}  
\begin{tabular}{cccccc} 

\hline \hline
Region & N0 & N1 & S0 & S1 & S2\\
\hline
$^a$ R$_m $ & 1.75 & 4.24 &1.75  & 4.01 & 7.35\\
  (kpc)   &       &       &     &      &    \\
\hline
[OIII] flux & 10.4 $\pm 0.2 $ & 34.5 $\pm 0.3 $ &  48.2 $\pm 0.4 $ &  41.6 $\pm 0.4 $& 49.9 $\pm 0.4  $\\
  ($10^{-15}$ erg/s/cm$^2$)& & & & \\
\hline
 X-ray flux (0.3-2.0 keV) & 2.9$^{+0.6}_{-0.7}$  &  8.3$^{+1.0}_{-0.8}$  &  20.2$^{+1.2}_{-1.3}$  &  10.4$^{+1.8}_{-1.5}$ & 7.8$^{+0.1}_{-2.5}$\\
 ($10^{-15}$ erg/s/cm$^2$)& & & & \\
\hline

ratio  & 3.6$^{+0.9}_{-0.8}$ &  4.2$^{+0.4}_{-0.5}$ &  2.4$^{+0.2 }_{-0.2}$ & 4.0$^{+0.6 }_{-0.7}$ & 6.4$^{+2.1}_{-0.1}$ \\
\hline \hline
\end{tabular}
\end{center}
\begin{flushleft}
\label{ratio_OX}
\tablecomments{\footnotesize{The flux of soft X-ray (0.3--2.0 keV) and the flux of [OIII] emission in each sub-cone region defined in Figure \ref{soft_ext} are listed. Also we calculate the flux ratio $f_{[OIII]}$/$f_{0.3-2.0 keV}$. $^a$ R$_m $ is the radii from the median of each zone to the nucleus of NGC 5252.}}

\end{flushleft}
 \end{table}  

 \clearpage

\begin{table}[tb]  
\caption{The fitting results of the photoionization model to the sub-cone spectra. }  
\begin{center}  
\begin{tabular}{cccccc}  
\hline \hline
 & $N_H$ & log$U$ & log$N_H$   & Flux & $\chi^2/d.o.f$ \\
 & ($\times 10^{22}$ cm$^{-2}$)  &         &   (cm$^{-2}$)  &($\times 10^{-15}$ erg/s/cm$^2$ ) & \\
\hline
 N0 & 0.22$^{+0.20}_{-0.12}$  & 0.14$^{+0.30}_{-0.24}$  & 20.6$^{+1.1}_{-1.6}$  & 2.9$^{+0.6}_{-0.7}$ & $10.87/13 $ \\
\hline
 N1 &0.21$^{+0.08}_{-0.09}$ &$-0.18^{+0.26}_{-0.21}$ & 20.4$^{+0.7}_{-0.5}$  &  8.3$^{+1.0}_{-0.8}$ & $ 17.81/19$ \\

\hline
 N2 & 0.34$^{+0.06}_{-0.05}$ & $-0.06^{+0.15}_{-0.12}$ & $21.5^{+0.4}_{-0.9}$ &  6.1$^{+0.8}_{-0.8}$ &$25.31
/24$ \\

\hline
 S0 &  0.0197* & $0.28^{+0.08}_{-0.06}$  &21.0$^{+0.5}_{-0.5}$  & 20.2$^{+1.2}_{-1.3}$ & $32.86/39 $\\

\hline
S1& 0.03$^{+0.06}_{-0.03}$& 0.05$^{+0.16}_{-0.18}$&  20.3$^{+0.4}_{-0.4}$ &  10.4$^{+1.8}_{-1.5}$ &$ 14.37/18$ \\

\hline
S2 &0.05$^{+0.17}_{-0.05}$ & $0.06^{+0.43}_{-0.46}$ &   19.0*  & 7.8$^{+0.1}_{-2.5}$ & $ 13.11/16$\\
   &0.15* & $-1.26$*         &   21.2*   &    &    \\

\hline
 S3 & $0.20^{+0.04}_{-0.06}$ & $-0.10^{+0.13}_{-0.12}$ &$21.8^{+0.3}_{-0.4}$  & 7.9$^{+0.9}_{-0.9}$ &$34.91/29 $ \\

\hline \hline
\end{tabular}
\end{center}
\tablecomments{\footnotesize{The spectra were extracted from each sub-cone region (shown in Figure \ref{soft_ext}). An absorbed photoionization model was applied to fit the spectra. The photoionization model is described in Section~\ref{sec:spec}. The  value of Galactic absorbing column towards NGC 5252, $N_H=0.0197\times 10^{22}$ cm$^{-2}$ is fixed for the minimum value of absorption. The other parameters with * are with unconstrained uncertainties and fixed to best fit values. A normalized nucleus component was added to fit the spectra extracted from S0, S1, N0 and N1, to account for the contribution from PSF. The nucleus contributes $\sim$ 0.8\% of its counts to S0 and N0 respectively, and $\sim$ 0.3\%--0.4\% to N1 and S1. The flux listed here is from the photoionized component.}}
\label{ionization}
\end{table}

\clearpage

\begin{table} 
\caption{Position angles of notable structures of NGC 5252.}  
\begin{center}  
\begin{tabular}{cc}  
\hline \hline   
PA of major axis of the soft X-ray extension & 170$^{\circ}$ \\ 
\hline  
PA of major axis of the host galaxy &   15$^{\circ}$ \\ 
\hline  
PA of major axis of [OIII] cone &   167$^{\circ}$  \\ 
\hline 
PA of the central band & 75$^{\circ}$  \\
\hline 
PA of the radio$-$emission & 170$^{\circ}$  \\
\hline \hline  
\end{tabular}
\end{center} 
\label{oriention} 
\begin{flushleft} 
\tablecomments{\footnotesize{The data, except the PA of the soft X-ray extension, are adopted from \citet{1989Natur.341..422T}. The ``central band'', defined as the linear region across the galactic nucleus and nearly perpendicular to the ionization cone, shows redder colour in the map of the ratio of [OIII] and the $H\alpha$. This phenomenon manifest lower ionization in the region, and may be resulted from obscuration along the central band \citep{1989Natur.341..422T}. }}
\end{flushleft}
\end{table}  

\clearpage

\begin{table} 
\caption{Fitting results of photoionization model for two side cones E and W}  
\begin{center}  
\begin{tabular}{ccccc} 

\hline \hline   
 & $N_H$ & log$U$ & log$N_H$  & $\chi^2/d.o.f$ \\
  & ($\times 10^{22}$ cm$^{-2}$)  &     &   (cm$^{-2}$)  & \\
\hline 

 E &   $ 0.41^{ +0.22}_{ -0.16}$ &   $ -0.94^{+0.73 }_{ -0.44}$  & 19  &  $ 5.99/20$  \\

\hline  
 W &  $ 0.36^{ +0.38}_{-0.20 }$ &   $-1.21 ^{ +0.85}_{-0.48 }$  & 19  &  $11.50/ 19$\\

\hline \hline  
\end{tabular}
\end{center}
\label{wholeph}
\begin{flushleft} 
\tablecomments{\footnotesize{The spectra were extracted from the two sectors east and west side of the nucleus, shown as the side cones E and W in Figure~\ref{sidecone}. They are modeled with the absorbed photoionized emission described in Section~\ref{sec:spec}. }
}
\end{flushleft}
 \end{table}  

 \clearpage







\bibliography{sample631}{}
\bibliographystyle{aasjournal}



\end{document}